\documentclass[twocolumn,nofootinbib,pra,preprintnumbers]{revtex4}
\usepackage{graphicx} %
\usepackage{subfigure}
\usepackage{hyperref}%

\usepackage{bm}
\usepackage{amsmath}
\usepackage{amssymb}
\usepackage{amsfonts}
\usepackage{fancyhdr}
\usepackage{appendix}
\usepackage{makeidx}

\begin{document}

\preprint{ ITP-UU-08/29, SPIN-08/22}

\title{Resolving Curvature Singularities in Holomorphic Gravity}

\author{Christiaan L.~M.~Mantz}
\email{CLMMantz@hotmail.com}
\author{Tomislav Prokopec}
\email{T.Prokopec@uu.nl} \affiliation{Institute for
Theoretical Physics, University of Utrecht, Princetonplein 5, P.O.
Box 80.006, 3508 TA Utrecht, The Netherlands}
\date{\textit{\today}}

\begin{abstract}
\noindent We formulate holomorphic theory of gravity
and study how the holomorphy symmetry alters the two
most important singular solutions of general relativity:
black holes and cosmology.
We show that typical observers (freely) falling into a holomorphic
black hole do not encounter a curvature singularity.
Likewise, typical observers do not experience Big Bang
singularity. Unlike Hermitian gravity~\cite{MantzHermitianGravity},
Holomorphic gravity does not respect the reciprocity symmetry and thus
it is mainly a toy model for a gravity theory formulated
on complex space-times. Yet it is a model that deserves a closer
investigation since in many aspects it resembles Hermitian gravity
and yet calculations are simpler.
We have indications that holomorphic gravity
reduces to the laws of general relativity correctly at large
distance scales.
\end{abstract}
%

\maketitle %

%
%
%
%
\pagenumbering{arabic}

\section{Introduction}
\noindent The existence of singularities in Einstein's well tested
theory of general relativity have puzzled many physicists since
their discovery. It is a goal of (and motivation for) theories as
quantum gravity to remove them. We propose generalizations of
general relativity which to a great extent ease the singularity problems
already at the classical level of the theory.
Singularities of general relativity are typically manifest as divergences
of some curvature invariant
and are ubiquitous in general relativity~\cite{Hawking:1973uf,Wald:1984rg}.

 Consider, for example, the Big Bang singularity.
A universe undergoing a power-law expansion expands with a scale factor,
$a\sim t^{1/\epsilon}$, where $\epsilon = -\dot H/H^2 = {\rm constant}\geq 0$
denotes the `slow roll' parameter, $H = \dot a/a$ is the Hubble expansion rate
and {\it dot} denotes a derivative with respect to physical time $t$.
The Ricci scalar $R$ then diverges as $t\rightarrow 0$
(corresponding to the {\it time} when matter density diverges),
\begin{equation}
 R = 6(2-\epsilon)H^2=\frac{6(2-\epsilon)}{\epsilon^2 t^2} \rightarrow \infty
\,,\qquad {\rm as} \;\; t\rightarrow 0
\,.
\label{BBsingularity}
\end{equation}
 Analogously, the Schwarzschild metric is singular when the coordinate radius
$r = \|\vec x \|\rightarrow 0$
(corresponding to the {\it place} where all of the mass is concentrated),
resulting in the curvature singularity of the Riemann tensor,
\begin{equation}
  R_{\mu\nu\rho\sigma}R^{\mu\nu\rho\sigma} = \frac{48G_N^2 M^2}{c^4r^6}
 \rightarrow \infty
\,,\qquad {\rm as} \;\; r\rightarrow 0
\,,
\label{BHsingularity}
\end{equation}
where $M$ denotes the black hole mass,
$G_N$ the Newton constant and $c$ the speed of light.

On complex manifolds space-time coordinates $x^\mu$
get complexified
as~\cite{MantzHermitianGravity},~\footnote{For the sake of simplicity
here we drop a factor $1/\sqrt{2}$ which we used
in Ref.~\cite{MantzHermitianGravity}
when relating $z^\mu$ to $x^\mu$ and $p^\mu$.}
\begin{equation}
 x^\mu \rightarrow z^\mu = x^\mu + i y^\mu
\,,\qquad y^\mu = \frac{G_N}{c^3}p^\mu
\,,
\label{complex coordinates}
\end{equation}
where $p^\mu$ denotes the energy-momentum four-vector of {\it noninertial
frames}, which can take any value; for an on-shell particle/observer however,
 $p^\mu$ reduces to the particle's
energy-momentum. Based on the holomorphy symmetry,
we expect that the singularities~(\ref{BBsingularity}--\ref{BHsingularity})
appear as some power of
\begin{eqnarray}\label{eq:one over z}
    \frac{1}{z},
\end{eqnarray}
(and possibly its complex conjugate), where
$z \rightarrow z^0 = x^0 + i y^0 = x^0 + i(G_N/c^4)E$
for the Big Bang singularity~(\ref{BBsingularity}) and
$z \rightarrow \|\vec z\,\| = \|\vec x\| + i \|\vec y\,\|
              = \|\vec x\| + i(G_N/c^3) \|\vec p\,\|$
for the black hole singularity~(\ref{BHsingularity}).
Near singularities the observed space-time curvature is typically
proportional to a power of the real part of~(\ref{eq:one over z}),
\begin{eqnarray}\nonumber
 \Re\bigg[\frac1z\bigg] =  \frac{x}{x^2+ y^2},
\end{eqnarray}
which blows up only when both $x$ and $y$ are {\it simultaneously} zero.
The aim is
now to formulate a generalization of general relativity that
yields such complexified singularities and show that for a freely
falling observer $x$ and $y$ are almost never zero
simultaneously.~\footnote{The meaning of `almost never' is made more precise below.}
In Hermitian gravity \cite{MantzHermitianGravity}, the Big Bang
singularity can be considered as `resolved' in the sense
that the set of observers moving backwards in time that
encounter the Big Bang singularity is of measure zero. So
typical observers do not see the singularity.
Note that this is \textit{not} the case in general relativity,
where all backward-moving observers eventually
hit the Bing Bang singularity. In this paper we present a new
theory, holomorphic gravity, in which also typical observers falling
into a Schwarzschild black hole do not encounter a singularity.
We believe that this resolution of singularities
is a generic feature of gravity theories
formulated on complex spaces as presented in this work and
in Ref.~\cite{MantzHermitianGravity},
representing one of the principal advantages of complex theories of gravity
when compared with Einstein's general relativity.

\section{Almost Complex Structure}
\noindent A natural generalization of general relativity, in order
to obtain complex solutions as (\ref{eq:one over z}), would be to
consider a theory with complex metrics, living on complex
manifolds. A general complex metric on a complex manifold is given
by
\begin{eqnarray}\label{complex metric}
   C
   =
   C_{\mu\nu} d z^{\mu} \otimes d z^{\nu}
   &+&
   C_{\mu\bar\nu} d z^{\mu} \otimes d z^{\bar\nu}
\\\nonumber
   +
   C_{\bar\mu\nu} d z^{\bar\mu} \otimes d z^{\nu}
   &+&
   C_{\bar\mu\bar\nu} d z^{\bar\mu} \otimes d z^{\bar\nu},
\end{eqnarray}
where barred indices $z^{\bar{\mu}} \equiv {\bar{z}}^{\mu}$ denote
complex conjugation. Hermitian gravity
\cite{MantzHermitianGravity} is formulated on a Hermitian manifold
endowed with a Hermitian metric, defined as follows
\begin{eqnarray}\label{eq:Hermitian metric formal definition}
    C_p \left( J_p Z ,  J_p W \right)
    =
     C_p \left( Z ,  W \right)
\,,
\end{eqnarray}
where the action of the almost complex structure operator, $J$, on
the basis vectors of the complexified tangent space is given by
\begin{eqnarray}\label{eq:almost complex structure z zbar operation}
    J_p\left(\frac{\partial}{\partial z^\mu}\right)
    =
    i \frac{\partial}{\partial z^\mu}
    \phantom{halloda}
    J_p\left(\frac{\partial}{\partial \bar{z}^\mu}\right)
    =
    - i \frac{\partial}{\partial \bar{z}^\mu}.
\end{eqnarray}
A Hermitian metric is a complex metric which has -- as a
consequence of the symmetry requirement~(\ref{eq:Hermitian metric formal
definition}) -- vanishing $C_{\mu\nu}$ and $C_{\bar\mu\bar\nu}$
components:
\begin{eqnarray}\nonumber
   C
   =
   C_{\mu\bar\nu} d z^{\mu} \otimes d z^{\bar\nu}
   +
   C_{\bar{\mu}\nu} d z^{\bar{\mu}} \otimes d z^{\nu}
\,.
\end{eqnarray}
The requirement that the complex metric satisfies Bianchi identities
lead us in~\cite{MantzHermitianGravity} to the conclusion
that Eq.~(\ref{eq:Hermitian metric formal definition})
can be consistently imposed on the metric only at the level
of the equations of motion (on-shell),
while at the level of the action (off-shell) all complex metrics of the
form~(\ref{complex metric}) are in fact allowed. Thus studying
holomorphic gravity allows one also to rectify the differences between
the full complex theory (which, apart from holomorphy on vielbeins,
has no additional symmetry requirements) and hermitian gravity.

One of the reasons why one wants to study a theory of gravity that
is invariant under the operation of $J$, is that the commutation
relations of quantum mechanics are invariant under its action,
which seems an invitation for quantization (recall that
the $y^\mu$ coordinate in~(\ref{complex coordinates})
is identified with the energy-momentum coordinate, $p^\mu$, in
the commutation relations).

We can also consider the theory which is anti-symmetric under the
action of the almost complex structure operator, in the sense that
the holomorphic metric is defined in the
following manner
\begin{eqnarray}\label{eq:holomorphic metric formal definition}
    C_p \left( J_p Z ,  J_p W \right)
    =
    - C_p \left( Z ,  W \right).
\end{eqnarray}
The holomorphic metric can then be written as
\begin{eqnarray}\label{eq:holomorphic metric}
   C
   =
   C_{\mu\nu} d z^{\mu} \otimes d z^{\nu}
   +
   C_{\bar{\mu}\bar{\nu}} d z^{\bar{\mu}} \otimes d z^{\bar{\nu}},
\end{eqnarray}
where the component $C_{\mu\nu}$ ($C_{\bar{\mu}\bar{\nu}}$ ) is
(anti-)holomorphic, which simplifies calculations. In the
remainder of this paper we construct a theory of gravity,
based on the holomorphic metric~(\ref{eq:holomorphic metric}).
In the flat space limit the holomorphic metric is invariant under
the complexified Lorentz group, $SO(1,3;\mathbb{C})$, while the line
element is invariant under the complexified inhomogeneous Lorentz group
(the complexified Poincar\'e group) $ISO(1,3;\mathbb{C})$.

\section{The Holomorphic Metric}

\noindent The holomorphic line element is defined in the following
manner
\begin{eqnarray}\label{eq:line element holomorphic metric familiar}
    d s^2
    =
    dz^{\mu}C_{\mu\nu}dz^{\nu}
    +
    dz^{\bar{\mu}}C_{\bar{\mu}\bar{\nu}}dz^{\bar{\nu}},
\end{eqnarray}
which can be written in eight dimensional notation
\begin{eqnarray}\nonumber
    ds^2
    =
    d\boldsymbol{z}^T\cdot \boldsymbol{C} \cdot d\boldsymbol{z}
    =
    (d\boldsymbol{z}^m)^T\boldsymbol{C}_{mn}d\boldsymbol{z}^n
    \\\label{eq:line element holomorphic metric subsection}
    =
       (dz^{\mu},
        dz^{\bar{\mu}})
    \left(\begin{array}{cc}
       C_{\mu\nu}& 0\\
        0 & C_{\bar{\mu}\bar{\nu}}
    \end{array}\right)
    \left(\begin{array}{c}
        dz^{\nu}\\
        dz^{\bar{\nu}}
    \end{array}\right),
\end{eqnarray}
where the Latin indices can take the values
$0,1,...,$$d-1$,$\bar{0},\bar{1},...,\overline{d-1}$, the
Greek indices run in the range $0,1,...,d-1$ and $d$
denotes the complex dimension of the complex manifold. The
entries of the metric $\bm{C}_{mn}$ are functions of holomorphic
and antiholomorphic vielbeins defined as
follows~\footnote{Here the Latin indices $a,b$ run from
$0,1,...,d-1$, since they represent local indices, and
$\eta_{\mu\nu} = $ diag$(-1,1,1,1)$.}
\begin{eqnarray}\label{eq:metric unrotated in terms of vielbeins}
    C_{\mu\nu}
    =
    e(z)^{\phantom{\mu}a}_{\mu}\eta_{ab}e(z)^{\phantom{\mu}b}_{\nu}
    \\\nonumber
     C_{\bar{\mu}\bar{\nu}}
    =
    e(\bar{z})^{\phantom{\bar{\bar{\mu}}}a}_{\mu}\eta_{ab}
          e(\bar{z})^{\phantom{\mu}b}_{\bar{\nu}}.
\end{eqnarray}
The metric components are symmetric under transposition, since
they are just an inner product of the vielbein times its
transpose. Hence also in eight dimensional notation the metric is
symmetric under transposition $\bm C^T = \bm C $. We define the
$z^\mu$ and ${{z}}^{\bar{\mu}}$ coordinates in terms of
$x^\mu$ and $y^{\check{\mu}}$, such that we obtain~\footnote{Checks
( $\check{}$ ) are put on indices to denote the imaginary part of
a coordinate or on indices of objects, which are projected onto
their basis vectors.}
\begin{eqnarray}\nonumber
    z^\mu
    =
    x^\mu+ i y^{\check{\mu}}\,,
    \phantom{hallo}
    \frac{\partial}{\partial z^\mu}
    &=&
    \frac{1}{2}\Big(\frac{\partial}{\partial x^\mu}
       - i \frac{\partial}{\partial y^{\check{\mu}}}\Big)
\end{eqnarray}
and their complex conjugates. This implies the following
decomposition of complex vielbeins in their real, ${e_R}^a_\mu$,
and imaginary, ${e_I}^a_{\check{\mu}}$, parts in the following
manner
\begin{eqnarray}\nonumber
    e_a^\mu
    &=&
    {e_R}_a^\mu + i {e_I}_a^{\check{\mu}}\,,
    \phantom{hallo}
    \overline{e_a^{\mu}}
    =
    e_a^{\bar{\mu}} = {e_R}_a^\mu - i{e_I}_a^{\check{\mu}}
    \\\label{eq:viebein decompositions}
    e^a_\mu
    &=&
    {e_R}^a_\mu - i {e_I}^a_{\check{\mu}}\,,
    \phantom{hallo}
    \overline{e^a_{\mu}}
    =
    e^a_{\bar{\mu}} = {e_R}^a_\mu + i{e_I}^a_{\check{\mu}}.
\end{eqnarray}
Vielbeins are holomorphic functions, and thus transform as
holomorphic vectors (we consider only the transformation of the
Greek indices for this purpose),
\begin{eqnarray}\label{tetrad:coord.transformations}
  &&e_\mu^a(z^\nu) \rightarrow
    \tilde e_\mu^a(w^\nu) = \frac{\partial z^\alpha(w^\nu)}
                                {\partial w^\mu}e_\alpha^a(z^\rho)
\\\nonumber
  &&e^\mu_a(z^\nu) \rightarrow
    \tilde e^\mu_b(w^\nu) = \frac{\partial w^\mu(z^\nu)}
                                 {\partial z^\alpha}e^\alpha_b(z^\rho)
\,.
\end{eqnarray}
The holomorphy of vielbeins,
\begin{eqnarray}\nonumber
    \frac{\partial}{\partial z^{\bar \mu}}e_\nu^a
    =
    \frac{1}{2}\Big[\frac{\partial}{\partial{x^{\mu}}}
                     + i\frac{\partial}{\partial y^{\check\mu}}\Big]
    [{e_R}_\nu^a + i {e_I}_{\check\nu}^a] =0
\,,
\end{eqnarray}
implies the Cauchy-Riemann equations,
\begin{equation}
  \frac{\partial{e_R}_\nu^a }{\partial x^{\mu}}
          = \frac{\partial{e_I}_{\check\nu}^a }{\partial y^{\check\mu}}
\,,\qquad
  \frac{\partial{e_I}_{\check\nu}^a }{\partial x^{\mu}}
         = -\frac{\partial{e_R}_\nu^a }{\partial y^{\check\mu}}
\,. \label{eq:CauchyRiemann}
\end{equation}
 When rotating the metric from $z$ and $\bar{z}$ space to $x$ and
 $y$ space, the components of the holomorphic rotated metric $\bm g_{mn}$ are
 given by
\begin{align}\label{eq:metric g holomorphic in C's}
    \bm g_{mn} =
    \left(\begin{array}{cc}
        g_{\mu\nu} & g_{\mu\check{\nu}} \\[.1cm]
        g_{\check{\mu}\nu} & g_{\check{\mu}\check{\nu}}
    \end{array}\right)
    =
    \frac{1}{2}
    \left(\begin{array}{cc}
        C_{\mu\nu}+C_{\bar{\mu}\bar{\nu}} & i(C_{\mu\nu} - C_{\bar{\mu}\bar{\nu}}) \\[.1cm]
        i(C_{\mu\nu} - C_{\bar{\mu}\bar{\nu}}) & - C_{\mu\nu} - C_{\bar{\mu}\bar{\nu}}
    \end{array}\right)\,.
\end{align}
Its inverse then becomes
\begin{align}\nonumber
    \bm g^{mn}
    =
    \frac{1}{2}
    \left(\begin{array}{cc}
        C^{\mu\nu} + C^{\bar{\mu}\bar{\nu}} & i(- C^{\mu\nu} + C^{\bar{\mu}\bar{\nu}})
\\[.1cm]
        i(- C^{\mu\nu} + C^{\bar{\mu}\bar{\nu}}) & -C^{\mu\nu} - C^{\bar{\mu}\bar{\nu}}
    \end{array}\right).
\end{align}
Clearly the entries of these rotated metrics are all symmetric and
real. It is easily verified that $C_{\mu\nu} = g_{\mu\nu} - iK_{\mu\nu}$
and that $C_{\bar{\mu}\bar{\nu}} = g_{\mu\nu}
+ i K_{\mu\nu}$, $K_{\mu\nu}=g_{\mu\check{\nu}}=g_{\check{\mu}\nu}$
is a symmetric metric tensor
and $g_{\check\mu\check\nu}=-g_{\mu\nu}$.
We can write the metric in terms of the
real and imaginary parts of the vielbein and in terms of the
imaginary part of $C_{\mu\nu}$, using the definition of the
complex metric in terms of vielbeins (\ref{eq:metric unrotated in
terms of vielbeins}), yields
\begin{eqnarray}\label{eq:metric g holomorphic in vielbeins}
       \boldsymbol{g}_{mn}
       =
         \left(\begin{array}{cc}
            e_{\mu} e_{\nu} - e_{\check{\mu}} e_{\check{\nu}} &  - e_{\check{\mu}} e_{\nu} - e_{\mu} e_{\check{\nu}}\\[.1cm]
            -e_{\check{\mu}} e_{\nu} - e_{\mu} e_{\check{\nu}} & - e_{\mu} e_{\nu} + e_{\check{\mu}} e_{\check{\nu}}
        \end{array}\right).
\end{eqnarray}
Expressing the inverse metric $\boldsymbol{g}^{mn} $ in terms of
vielbeins yields
\begin{align}\label{eq:rotated metric from z to x space in vielbeins}
        \left(\begin{array}{cc}
            g^{\mu\nu}          &   K^{\mu\nu}\\[.1cm]
            K^{\mu\nu}  &   -g^{\mu\nu}
        \end{array}\right)
        =
    \left(\begin{array}{cc}
        e^{\mu} e^{\nu} - e^{\check{\mu}} e^{\check{\nu}} &   e^{\check{\mu}} e^{\nu} + e^{\mu} e^{\check{\nu}}\\[.1cm]
        e^{\check{\mu}} e^{\nu} + e^{\mu} e^{\check{\nu}} & - e^{\mu} e^{\nu} + e^{\check{\mu}} e^{\check{\nu}}
    \end{array}\right).
\end{align}
With the rotated metric we can write down the holomorphic line
element in its rotated form
\begin{align}\label{eq:holomorphic line element rotated}
    d s^2
    =
    g_{\mu\nu}d x^\mu d x^\nu
    -
    \frac{2 G_N}{c^3} K_{\mu\nu}d p^{\mu} d x^\nu
    -
   \frac{ G_N^2}{c^6} g_{\mu\nu}d p^{\mu} d p^{\nu}\,.
\end{align}
From the
definition of the complex metric in terms of vielbeins
(\ref{eq:metric unrotated in terms of vielbeins}) and the fact
that the vielbeins are defined to satisfy
\begin{eqnarray}\nonumber
    e^\mu_a e^a_\nu
    =
    \delta^\mu_\nu ,
\end{eqnarray}
it follows that $ \bm g^{me}$ and $\bm g_{en}$ are inverses of
each other
\begin{eqnarray}\nonumber
    \bm g^{me}\bm g_{en}
    =
    \bm\delta^m_n ,
\end{eqnarray}
where $\bm\delta^m_n $ is defined as
\begin{eqnarray}\nonumber
    \bm\delta^m_n
    \equiv
    \left(
    \begin{array}{cc}
        \delta^\mu_\nu  &           0 \\
        0               &           \delta^{\bar{\mu}}_{\bar{\nu}}
    \end{array}
    \right).
\end{eqnarray}

\section{Flat Space}

\noindent The holomorphic line element~(\ref{eq:holomorphic line element rotated})
in flat space becomes
\begin{eqnarray}
    ds^2
    =
    -(c d t)^2 + (d\vec{x})^2 -[-(d y^0)^2 + (d\vec{y})^2 ]
\,.
\label{line-element:0}
\end{eqnarray}
Recall that the $y$ coordinate is related to the
energy-momentum coordinate~\cite{MantzHermitianGravity,Low:2006qm} as
$y^\mu \equiv p^\mu{G_N}/{c^3}$.  The space-time-momentum-energy interval
squared from the origin to a space-time-momentum-energy point
${\bm x}^m$ is given by
\begin{align}\label{eq:line element flat space holomorphic}
    d^2\left(\bm 0;{\bm x}^m\right)
    = -(ct)^2 + (\vec{x})^2
    - \frac{G_N^2}{c^6}\left[-\left(\frac{E}{c}\right)^2 + (\vec{p})^2\right]
\,.
\end{align}
Note that the contribution of momentum-energy
to the line element multiplies
${G_N^2}/{c^6} $, which is tiny (${G_N^2}/{c^6}\sim 10^{-72}{\rm s^2/kg^2}$),
as it should be, since one does not observe
any momentum-energy contribution to~(\ref{eq:line element flat space holomorphic})
at low energies. For light-like propagation ($ds^2 = 0$)
and in the absence of momentum-energy contribution,
Eq.~(\ref{eq:line element flat space holomorphic}) reduces to the well known
result: massless particles move on the light-cone with the speed of light,
$v = \|d\vec{x}/dt\| =c $.

\begin{figure}
  \centering
  \includegraphics[width= 7cm]{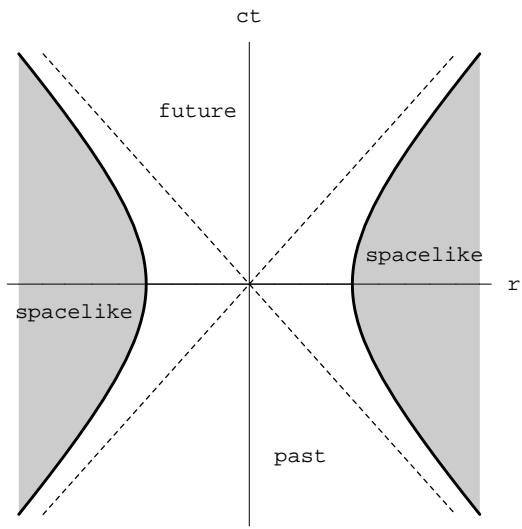}
  \caption{The figure shows how the light cones get modified
  by non inertial coordinate transformations
  on a space-time-momentum-energy diagram in the case when $p^2>0$.
  Note that there is a non-local instantaneous causally
  related volume element near the origin.}
  \label{fig:LightConehol1}
\end{figure}

On the other hand, for a light-like propagation
in the presence of a non-vanishing momentum-energy contributions however,
Eq.~(\ref{eq:line element flat space holomorphic}) yields,
\begin{eqnarray}\label{eq:distance flatspace hol on-shell}
-(ct)^2 + (\vec{x})^2 - \frac{p^2G_N^2}{c^6} = 0,
\end{eqnarray}
where the variable $p^2=\eta_{\mu\nu}p^\mu p^\nu$ is the
energy-momentum squared in a space-time-momentum-energy hyper-surface.
Setting the space-time-momentum-energy interval to zero determines the
boundary of causality (light cones).
We can specify our hyper-surface
further by setting $t$ to zero. Assuming
$p^2>0$, one finds that there is a spatial region
which is in instantaneous causal contact, and whose radius is given by,
\begin{eqnarray}\nonumber
    r_{max}
    =
    \frac{\sqrt{p^2} G_N}{c^3}\qquad (p^2>0)
\,.
\end{eqnarray}
The causally related regions (time-like, $ds^2<0$) for the case
$p^2>0$ are shown as white in figure~\ref{fig:LightConehol1},
where the region of non-local causal contact at $t=0$ is clearly seen
as the hyperbola's throat.
\begin{figure}
  \centering
  \includegraphics[width= 7cm]{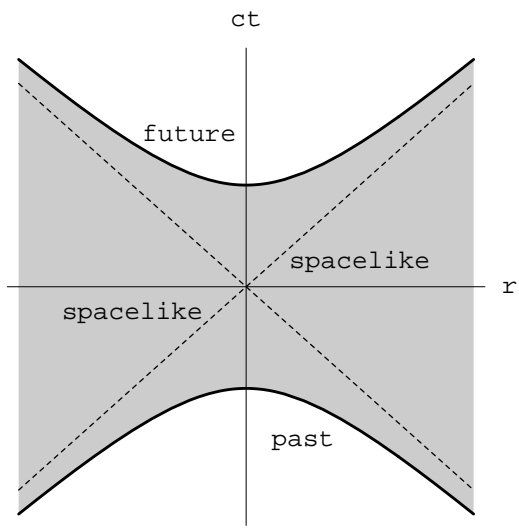}
  \caption{The figure shows how the light cones get modified
  by non-inertial coordinate transformations when $p^2<0$.
  There is a minimum time interval required
  for events to be in causal contact.}
  \label{fig:LightConeHol}
\end{figure}
On the other hand, when $p^2<0$, the causal boundaries, shown in
figure~\ref{fig:LightConeHol}, are quite different. There
is a minimal time $t_{min}$ required for events to be in causal contact,
\begin{eqnarray}\nonumber
    t_{min}
    =
    \frac{\sqrt{- p^2} G_N}{c^4}\qquad (p^2<0)
\,.
\end{eqnarray}

We now calculate the phase velocity, using the
space-time-momentum-energy line element (\ref{eq:distance
flatspace hol on-shell})
\begin{eqnarray}\label{eq:phase velocity hermitian case}
    v_{\rm phase}
    =
    \frac{x}{t}
    = \sqrt{c^2  + \frac{p^2G_N^2}{c^4t^2}}.
\end{eqnarray}
One can see that the phase velocity 
approaches the speed of light for large $t$.

On the other hand, the group velocity becomes
\begin{eqnarray}\label{eq:groupvelocity holomorphic case}
 v_{g} = \sqrt{c^2 + \frac{G_N^2}{c^6}f^2}
\,,
\end{eqnarray}
where
\begin{eqnarray}\nonumber
    f^2
    =
 -\frac{1}{c^2}\left(\frac{dE}{dt}\right)^2
 +\left(\frac{d\vec{p}}{dt}\right)^2
\,.
\end{eqnarray}
The group velocity approaches the speed of light for small four-forces
squared. Requiring that $v_g$ be real and positive
(as required by propagation) sets a lower limit on the four force,
$f^2\geq -c^8/G_N^2$. This is to be contrasted with an upper limit
occurring in Hermitian gravity~\cite{MantzHermitianGravity}.
Just like in the case of Hermitian gravity, we expect that these
aparent violations of causality become of importance only
in the regions where the particle's self-gravity is large
(in the vicinity of particle's own event horizon).

\section{Second Order Formalism}

\noindent

Let us now state the holomorphic Einstein-Hilbert action,~\footnote{In contrast
to Hermitian gravity~\cite{{MantzHermitianGravity}}, the holomorphy symmetry
of the metric tensor of holomorphic gravity can be imposed both
on- and off-shell (at the level of the action). This has the advantage that
the action~(\ref{eq:holomorphic Einstein-Hilbert action})
suffices to fully specify the dynamics of
holomorphic gravity, and no further (action) constraints are needed.
Furthermore, it can be easily shown that the Bianchi identities and
the covariant stress-energy conservation are satisfied.}
\begin{eqnarray}\label{eq:holomorphic Einstein-Hilbert action}
    S
    =
     \frac{c^4}{16\pi G_N}\int d^8z\sqrt{-C} R  + {\rm c.c.}
\end{eqnarray}
The holomorphic Ricci scalar
is defined by $R = C^{\mu\nu}R_{\mu\nu}$, where the
holomorphic Ricci tensor is given by
\begin{eqnarray}\label{eq:holomorphic Ricci tensor}
    R^{\lambda}_{\mu\lambda\nu}
        =
        \partial_{\lambda}\Gamma^{\lambda}_{\nu\mu} -
        \partial_{\nu}\Gamma^{\lambda}_{\lambda\mu}
        +
        \Gamma^{\lambda}_{\lambda\alpha}\Gamma^{\alpha}_{\nu\mu} -
        \Gamma^{\lambda}_{\nu\alpha}\Gamma^{\alpha}_{\lambda\mu}
\,,
\end{eqnarray}
and where here $\partial_\lambda = {\partial}/{\partial z^\lambda}$.
Since the holomorphic Einstein-Hilbert action is just
the complexified equivalent of the Einstein-Hilbert action of standard general relativity,
the equations of motions for this theory are just complexified
equivalents of the equations of motions of general relativity.
Using first order formalism \cite{MantzThesis:2007},
the equations of motion implied by the
action~(\ref{eq:holomorphic Einstein-Hilbert action}) are:
\begin{subequations}
\begin{eqnarray}\label{eq:holomorphic Einstein equations}
    G_{\mu\nu}
    &=&
   \frac{ 8\pi G_N}{c^4} T_{\mu\nu}\\
\label{eq:holomorphic metric compatibility equations}
   \nabla_\rho C^{\mu\nu} &=& 0
\,.
\end{eqnarray}
\end{subequations}
Through the holomorphic metric compatibility equations
(\ref{eq:holomorphic metric compatibility equations}) the
holomorphic connection coefficients are given by
\begin{eqnarray}\label{eq:holomorphic Levi-Civita
connection}
    \Gamma_{\mu\nu}^{\rho}
    =
    \frac{1}{2}C^{\rho\epsilon}(\partial_{\mu}C_{\epsilon\nu}
    + \partial_{\nu}C_{\mu\epsilon} -
    \partial_{\epsilon}C_{\mu\nu})
\,.
\end{eqnarray}
Plugging the holomorphic Levi-Civit\`a connection coefficients into the
holomorphic Einstein's equations (\ref{eq:holomorphic Einstein
equations}), we obtain second order differential equations in
terms of the holomorphic metric only.

When varying the holomorphic line element (\ref{eq:line element
holomorphic metric subsection}) we obtain the following
holomorphic geodesic equations
\begin{eqnarray}\label{eq:holomorphic geodesic equations}
    \ddot{z}^\rho
    +
    \Gamma_{\mu\nu}^{\rho}\dot{z}^{\mu}\dot{z}^{\nu}
    =
    0
\end{eqnarray}
and its complex conjugate. We can write
equations~(\ref{eq:holomorphic geodesic equations})
in their convenient eight dimensional form, such that we
can rotate them to $x,y$ space, by simply plugging in the rotated
metric (\ref{eq:rotated metric from z to x space in vielbeins}).
This is possible because the holomorphic connection coefficients
transform as a (1,2) tensor under the constant coordinate
transformations such as rotations from $z,\bar{z}$ to $x,y$ space. The
rotated eight dimensional connection coefficients are then simply given
by
\begin{eqnarray}\label{eq:connection coefficients etries arbitrary x,y written
out}
    {\bm\Gamma}_{mn}^{r}
    =
    \frac{1}{2}{\bm g}^{re}(\partial_{m}{\bm g}_{en}
    + \partial_{n}{\bm g}_{me} -
    \partial_{e}{\bm g}_{mn})
\,,
\end{eqnarray}
where the metric $\bm g_{mn}$ is just the eight dimensional rotated
metric (\ref{eq:metric g holomorphic in C's}).

\section{The Limit to General Relativity}
\noindent The limit of Holomorphic gravity to general relativity
is based on the assumption that the $y$ coordinate and its
corresponding vielbein are small. When expanding the equations of
holomorphic gravity in powers of $y$ and its corresponding
vielbein, we hope to obtain the theory of general relativity
at zeroth order of the expansion and meaningful corrections at
linear order.
The easiest way to obtain the limit to general relativity is to expand
the vielbeins in terms of the
$y$ coordinate
\begin{equation}
  e_\mu^a(x,y) = {\rm e}^{iy\cdot \partial_x}e_\mu^a(x)
\,.
\label{vielbein:expansion}
\end{equation}
From this and the definition, $C_{\mu\nu}=e_\mu(z)\cdot e_\nu(z)$,
we conclude that the analogous formula holds for the holomorphic
metric tensor,
$C_{\mu\nu}(x,y) = {\rm e}^{iy\cdot \partial_x}C_{\mu\nu}(x)$,
such that,
\begin{eqnarray}
 g_{\mu\nu}(x,y)   &=&\;\;\; \cos\left(y\cdot \partial_x\right)g_{\mu\nu}(x)
                   + \sin\left(y\cdot \partial_x\right)K_{\mu\nu}(x)
\nonumber\\\nonumber
 K_{\mu\nu}(x,y)   &=& -\sin\left(y\cdot \partial_x\right)g_{\mu\nu}(x)
                   + \cos\left(y\cdot \partial_x\right)K_{\mu\nu}(x)
\,,
\\\label{gK:expand}
\end{eqnarray}
where we made use of Eqs.~(\ref{eq:metric g holomorphic in C's})
and~(\ref{eq:rotated metric from z to x space in vielbeins}). The
analogous relations hold for the inverses $g^{\mu\nu}$ and
$K^{\mu\nu}$. Assuming that $K_{\mu\nu}$ is -- just like $y^\mu$
-- a first order quantity, Eq.~(\ref{gK:expand}) then tells us
that $g_{\mu\nu}(x,y)$ acquires corrections at second order in
$y$, while $K_{\mu\nu}(x,y)$ acquires first order corrections in
$y$, with $K_{\mu\nu}$ being on its own a first order quantity.

This analysis then implies that the connection with all
indices unchecked yields the ordinary Levi-Civit\`a connection
plus corrections of second order
\begin{eqnarray}\nonumber
   \Gamma ^\rho_{\mu\nu} (x,y) =  \Gamma ^\rho_{\mu\nu} (x) +
   O(y^2).
\end{eqnarray}
 With these connection coefficients, we can know check if the
theory reduces to the theory of general relativity by plugging
them into the rotated eight dimensional geodesic equation. Keeping
only terms of linear order in the $y$ coordinate and its
corresponding vielbein, yields the ordinary geodesic equation
\begin{eqnarray}\nonumber
    \ddot{x}^\rho + \Gamma_{\mu\nu}^\rho(x){\dot{x}}^\mu {\dot{x}}^\nu
    + O(y^2) = 0
\end{eqnarray}
without any first order corrections present.
We have not studied in detail the reduction 
of holomorphic Einstein's equations
to general relativity. Our study of holomophic
Schwarzschild solution and cosmology 
in sections~\ref{The holomorphic Schwarzschild Solution}
and~\ref{Cosmology} suggests however
that holomorphic gravity reduces to 
Einstein's theory in the low energy limit. 

\subsection{Scalar Field}

\noindent In this subsection we consider the following scalar
field action
\begin{eqnarray}\label{eq:scalar field action Hermitian}
    S_\phi
    =
    \int d^8 z \sqrt{- C}
    \left[- \frac{1}{2}C^{\mu\nu} (\nabla_{\mu}\phi)\nabla_\nu\phi - V(\phi)\right]
    + {\rm c.c.}
\,,
\end{eqnarray}
where $\phi=\phi(z^\mu)$ is a holomorphic function of $z^\mu$.
The equations of motion for the scalar field are then
\begin{eqnarray}\label{eq:scalar field holomorphic equations of motion}
    \square\phi - \frac{d V}{d\phi}
    =
    0
\end{eqnarray}
and its complex conjugate, where $\square =
C^{\mu\nu}\nabla_\mu\nabla_\nu$ and $\overline{\square} =
C^{\bar{\mu}\bar{\nu}}\nabla_{\bar{\mu}}\nabla_{\bar{\nu}}$. When
varying this action with respect to the metric we obtain the two
sets of components of the stress-energy tensor
\begin{eqnarray}\label{eq:stres-energy tensor holomorphic}
    T_{\alpha\beta}
    =
     \nabla_{\alpha}\phi\nabla_\beta\phi
    -\frac{1}{2}C_{\alpha\beta}
    C^{\mu\nu} \nabla_{\mu}\phi\nabla_\nu\phi
    - C_{\alpha\beta} V(\phi) 
\end{eqnarray}
and its complex conjugate. The energy conservation equations are
\begin{eqnarray}\nonumber
    \nabla_{\mu}T^{\mu\nu}
    =
    \left(\square \phi - \frac{d V}{d \phi}\right)\nabla^\nu \phi
    =
    0
\end{eqnarray}
and its complex conjugate, where we have used
Eq.~(\ref{eq:scalar field holomorphic equations of motion}).
There are two sets of Einstein's equations, namely
\begin{eqnarray}\label{eq:Einstein eq holomorphic scalar field}
    R_{\mu\nu}
    =
    \frac{8 \pi G_N}{c^4}
     \left(\nabla_{\mu} \phi\nabla_{\nu}\phi + C_{\mu\nu} V(\phi)\right)
\end{eqnarray}
and again its complex conjugate. The energy-momentum tensor
of a perfect fluid in the fluid rest frame
can be written in the diagonal form
\begin{eqnarray}\label{eq:energy momentum holomorphic tens invar rho and p}
    T^\mu_{\phantom{\mu}\nu} = \text{diag}(-\rho,p,p,p)
\end{eqnarray}
plus its complex conjugate, where $\rho$ is the density and $p$ the
pressure. For a homogeneous scalar field $\phi=\phi(t)$,
Eq.~(\ref{eq:stres-energy tensor holomorphic}) yields,
$\rho=(1/2c^2)\dot\phi^2+V(\phi)$ and $p=(1/2c^2)\dot\phi^2-V(\phi)$.

\section{The holomorphic Schwarzschild Solution}
\label{The holomorphic Schwarzschild Solution}

\noindent The holomorphic Schwarzschild solution should just be  a
complexified version of the ordinary Schwarzschild solution, since
the holomorphic Einstein's equations (\ref{eq:holomorphic Einstein
equations}) are just complexified versions of the ordinary
Einstein's equations. So we expect to have the holomorphic metric
components
\begin{eqnarray}\nonumber
    C_{00}
    =
    -\left(1 - \frac{1}{c^2}\frac{2 G_N M}{z}\right)\,,
    \\\label{eq:Schwarzschild coeff unrotated }
    C_{11}
    =
    \left(1 - \frac{1}{c^2}\frac{2 G_N M}{z}\right)^{-1}
\end{eqnarray}
and their complex conjugates, where $z = \|\vec{z}\|
=  \|\vec{x}\| + i\frac{G_N}{c^3} \|\vec{p}\|
\equiv r + i\frac{G_N}{c^3}p$
and $M$ is the mass of the black hole, which in holomorphic gravity
could, in principle, be
complex.~\footnote{At the moment we do not have a good physical
interpretation for $\Im[M]$.}
The angular part takes the same form as in general
relativity
\begin{eqnarray}\label{eq:angular part unrotated}
    d\Omega^2
    =
    d\theta^2 + \text{sin}^2\theta d\phi^2,
\end{eqnarray}
but now the spherical symmetry is $O(3;\mathbb{C})$
and thus the angular parameters are complex
\begin{eqnarray}\nonumber
     \theta
     =
     {\theta_R}
     +
     i {\theta_I}
\phantom{hallo}
     \phi
     =
     {\phi_R}
     +
     i {\phi_I}.
\end{eqnarray}
Thus the angular part of the holomorphic Schwarzschild solutions is
the complex angular part (\ref{eq:angular part unrotated}) plus
its complex conjugate. One can easily check that the holomorphic
Schwarzschild solution is indeed a solution to the holomorphic
Einstein's equations (\ref{eq:holomorphic Einstein equations}) in
vacu\"um. Expressing the Schwarzschild components
(\ref{eq:Schwarzschild coeff unrotated }) in $r$ and $p$ yields
\begin{eqnarray}\nonumber
    C_{00}
    &=&
    -\left(1- \frac{2 G_N M}{c^2}
     \frac{r - i \frac{G_N}{c^3}p}{r^2 + \frac{G_N^2}{c^6} p^2}\right)
\\\label{eq:holomorphic schwarz metric components}
    C_{11}
    &=&
    \frac{r^2 + \frac{G_N^2}{c^6} p^2 - \frac{2 G_N M}{c^2}
   (r + i \frac{G_N}{c^3}p) }
   {(r - \frac{2 G_N M}{c^2})^2 + \frac{G_N^2}{c^6}  p^2}
     = -\frac{1}{C_{00}}
\end{eqnarray}
and their complex conjugates. The rotated holomorphic metric
components (\ref{eq:metric g holomorphic in C's}) are given in
terms of the holomorphic metric components (\ref{eq:holomorphic
schwarz metric components}), and take the following form
\begin{eqnarray}\nonumber
    g_{00}
    &=&
    -\left(1 - \frac{2 G_N M}{c^2}
           \frac{r}{r^2 + \frac{G_N^2}{c^6} p^2}
     \right)
    \\\nonumber
    g_{11}
    &=&
    \frac{r^2 + \frac{G_N^2}{c^6} p^2
        - \frac{2 G_N M}{c^2} r  }{(r - \frac{2 G_N M}{c^2})^2 + \frac{G_N^2}{c^6}  p^2}
    \\\nonumber
    g_{0\check{0}}
    &=&
   \frac{ \frac{2 G_N^2 M}{c^5}p}{r^2 + \frac{G_N^2}{c^6} p^2}
    \\\nonumber
    g_{1\check{1}}
    &=&
   \frac{ \frac{2 G_N^2 M}{c^5}p }{(r - \frac{2 G_N M}{c^2})^2 + \frac{G_N^2}{c^6}  p^2}
\,,
\end{eqnarray}
which reduces to the Schwarzschild solution of general relativity when $p\rightarrow 0$.
When these rotated components are inserted into the rotated holomorphic
line element~(\ref{eq:holomorphic line element rotated}),
one obtains the holomorphic Schwarzschild metric in its full glory
\begin{align}\nonumber
   &d s^2
   =
   \\\nonumber
   &-\left(1- \frac{2 G_N M}{c^2} \frac{r}{r^2 + \frac{G_N^2}{c^6} p^2 }\right)
   \left[(c d{t})^2 - \frac{G_N^2}{c^8}\left(  d E \right)^2
   \right]\\\nonumber
   &+
   \frac{r^2 + \frac{G_N^2}{c^6} p^2
    - \frac{2 G_N M}{c^2} r }{(r - \frac{2 G_N M}{c^2})^2 + \frac{G_N^2}{c^6} p^2}
   \left[(d{r})^2 - \frac{G_N^2}{c^6}(d{p})^2\right]
  \\\nonumber
   &+
   \frac{ \frac{2G_N^2M}{c^5}p }
        {(r - \frac{2 G_N M}{c^2})^2 + \frac{G_N^2}{c^6}  p^2}
   \frac{2 G_N}{c^3}d r d p
    \\\nonumber
   &+
   \frac{ \frac{2 G_N^2 M}{c^5}p}{r^2 + \frac{G_N^2}{c^6}
   p^2}\frac{2 G_N}{c^4}d t d E
   +
   \left(r^2 - \frac{G_N^2}{c^6} p^2\right)\left(d{\theta_R}^2 -d{\theta_I}^2\right)
   \\\nonumber
   &-4 \frac{G_N}{c^3} r p d{\theta_R} d{\theta_I}
   +\left(d{\phi_R}^2 -d{\phi_I}^2\right)\times
   \\\nonumber
   &\times\Big[\left(r^2 - \frac{G_N^2}{c^6}p^2\right)\{(\text{sin}{\theta_R}\text{cosh}{\theta_I})^2 -(\text{cos}{\theta_R}\text{sinh}{\theta_I})^2 \}
   \\\nonumber
   &-4\frac{G_N}{c^3}r p (\text{sin}{\theta_R}\text{cosh}{\theta_I}\text{cos}{\theta_R}\text{sinh}{\theta_I})\Big]
   \\\nonumber
   &-2d{\phi_R}d{\phi_I}\Big[ \left(r^2 - \frac{G_N^2}{c^6}p^2\right)2\text{sin}{\theta_R}\text{cosh}{\theta_I}\text{cos}{\theta_R}\text{sinh}{\theta_I}
   \\\nonumber
   &+ 2\frac{G_N}{c^3} r p \{(\text{sin}{\theta_R}\text{cosh}{\theta_I})^2 -(\text{cos}{\theta_R}\text{sinh}{\theta_I})^2 \} \Big]
   .
\end{align}
This rotated holomorphic Schwarzschild metric is indeed
a solution of the rotated holomorphic Einstein's equations in
vacu\"um, as can be checked by explicit calculation of
$R_{\mu\nu}$. It is easy to see that, in the limit when the radius
$r$ goes to infinity and $p$ to zero, the solution approaches the
Minkowski metric, whereas the solution approaches the momentum-energy
Minkowski space when $r$ goes to zero and $p$ goes to
infinity. If $p$ is not zero when $r$ is zero, there is no
curvature singularity at the origin. Explicit calculation shows that
$p$ is not zero when $r$ is zero for a generic infalling observer. When
considering an observer which is falling in radially we can
neglect the change in angular coordinates,
corresponding to a vanishing angular momentum.
(Recall that in general relativity this choice
corresponds to the worst case scenario, for which
an observer descends in the quickest possible way towards
the black hole singularity.)
Based on the holomorphic Schwarzschild
metric~(\ref{eq:Schwarzschild coeff unrotated })
the black hole line element~(\ref{eq:line element holomorphic metric familiar})
of a radially infalling observer can be recast as
\begin{align}\nonumber
    -2c^2
    =
    -&\left(1 - \frac{1}{c^2}\frac{2 G_N M}{z}\right)
              \left(\frac{d z^0}{d \tau}\right)^2
    \\\label{eq:schwz unrotated para}
    +
    &\left(1
    - \frac{1}{c^2}\frac{2 G_N M}{z}\right)^{-1}\left(\frac{d z}{d \tau}\right)^2
    + \text{c.c,}\,,
\end{align}
where we introduced a real affine parameter (proper time) ${\tau }$ defined as
$c^2(d\tau)^2 = -ds^2$.
The holomorphic Schwarzschild solution has an isometry in the $z^0$ direction
($\partial/\partial z^0$ is a Killing vector)
 and hence there is a conserved quantity:
\begin{equation}
 e_z = u^0_0 + i\frac{G_N}{c^3}f^0_0
\,.
\label{e_z}
\end{equation}
The real part $u^0_0 = u^0(\tau_0)$ corresponds to the (initial) zeroth component of
the four-velocity evaluated at $\tau=\tau_0$
(the energy per unit mass divided by $c$), which is
also the Killing vector in general relativity,
while the imaginary part $f^0_0=f^0(\tau_0) = (dp^0/d\tau)(\tau_0)$
is the zeroth component of the observer's four-force.
Since $f^0_0$ does not contribute in general relativity,
the general relativistic limit should be obtained by setting $f^0_0\rightarrow 0$,
which is indeed the case.
In analogy with general relativity, the conserved quantity
can be written as~\cite{Wald:1984rg}
\begin{align}\label{eq:conserved killing quantity}
    \left(\frac{d z^0}{d \tau}\right)^2
    =
    \left(1 - \frac{1}{c^2}\frac{2 G_N M}{z}\right)^{-2}e_z^2\,.
\end{align}

 Equations~(\ref{eq:schwz unrotated para}) and~(\ref{eq:conserved killing quantity})
are not enough to fully determine $z=z(\tau)$, because nothing is known
about the imaginary part of the expressions appearing
in Eq.~(\ref{eq:schwz unrotated para}).
Since we are in holomorphic gravity, it is reasonable to demand $z=z(\tau)$
to be a holomorphic function. With this {\it analytic extension}
we can now completely determine $z=z(\tau)$
by inserting the conserved integral~(\ref{eq:conserved killing quantity})
into~(\ref{eq:schwz unrotated para}), to obtain
\begin{align}\label{eq:numerical}
   \left(\frac{d z}{cd \tau}\right)^2
    =
    b + \frac{r_s}{z}
\,,\qquad b=\frac{e_z^2}{c^2}-1
\,,\qquad r_s=\frac{2G_NM}{c^2}
\end{align}
and the complex conjugate of this equation.
Note that $r_s$ is just the Schwarzschild radius and the on-shell
value of $b=b_r+i b_i$ can be expressed in terms of the initial 3-velocity,
$b_r = \gamma_0^2 -1 = (\gamma_0 \vec v_0/c)^2$,
$\gamma_0^{-2} = 1-(\vec v_0/c)^2$.
Solving~(\ref{eq:numerical}) for proper time $\tau$ yields
\begin{align}\nonumber
    \frac{c\tau}{r_s}
    &=
    \frac{1}{b}\sqrt{\zeta(1+b\zeta)}
    \\\nonumber
    &-\frac{1}{b^{\frac{3}{2}}}
   \text{ln}\left\{\sqrt{b\zeta}
   +
   \sqrt{1+b\zeta}\right\}
\,,\quad \zeta = \frac{z}{r_s}
\,.
\end{align}
From this it follows that, when $b$ is complex and $r\rightarrow 0$,
$p$ is not zero. Moreover, one can show that quite generically,
when $r\rightarrow 0$, $p$ grows large, which limits the growth
of the curvature invariant~(\ref{BHsingularity}),
which for Holomorphic gravity has the simple generalization,
\begin{equation}
  R_{\mu\nu\rho\sigma}R^{\mu\nu\rho\sigma} = \frac{12 r_s^2}{z^6}
\,.
\label{BHsingularity:hol}
\end{equation}
For example, when $r\rightarrow 0$, this reduces to
$R_{\mu\nu\rho\sigma}R^{\mu\nu\rho\sigma} \rightarrow -12 c^{18}r_s^2/(G_Np)^6$,
which is negative.

\begin{figure}[ht]
  \centering
    \subfigure[]{
        \includegraphics[width=\columnwidth]{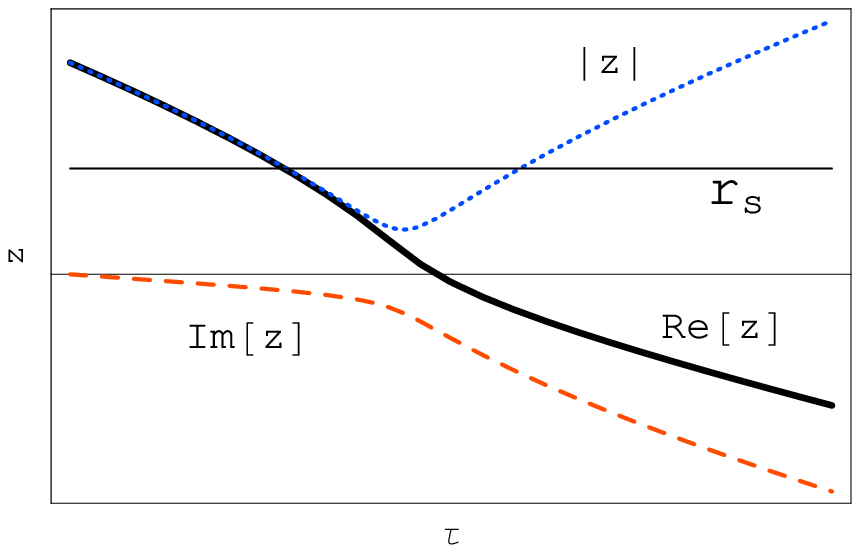}
        \label{fig:Radius1}
    }
    \subfigure[]{
        \includegraphics[width=\columnwidth]{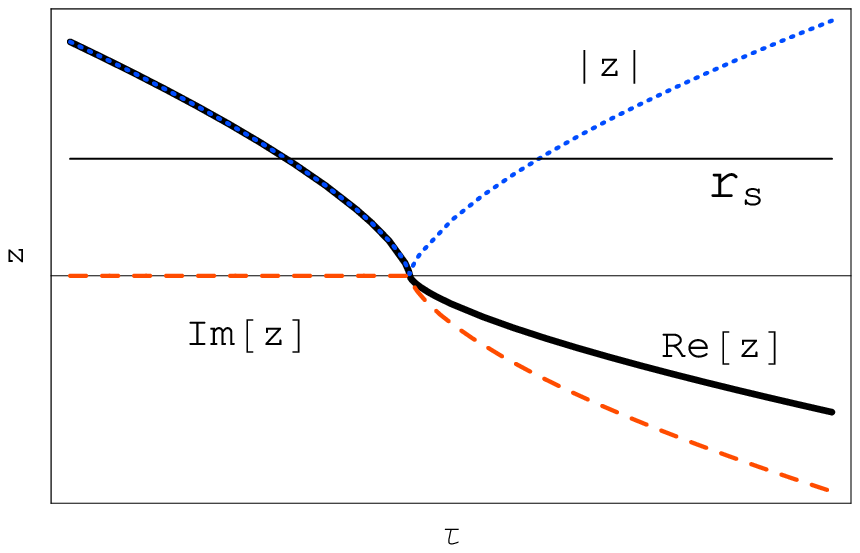}
        \label{fig:Radius2}
    }
    \caption{The real and imaginary part and the absolute value of the
    complex radial coordinate $z$
     as a function of proper time $\tau$ for a freely falling observer
     for different initial conditions $b$.
     \ref{fig:Radius1} The real and imaginary part and the absolute value of $z$
      with a complex initial condition $b$. In this case $|z|=0$
     is never reached, indicating
     that the observer encounters no black hole singularity.
    \ref{fig:Radius2} The real and imaginary part and the absolute value of $z$
    for a real initial $b$. In this case both $\Re[z]$ and $\Im[z]$ reach simultaneously
    {\it zero} for some value of proper time $\tau$, indicating a black hole singularity.}
    \label{fig:Radius}
\end{figure}

\begin{figure}[ht]
  \centering
    \subfigure[]{
        \includegraphics[width=\columnwidth]{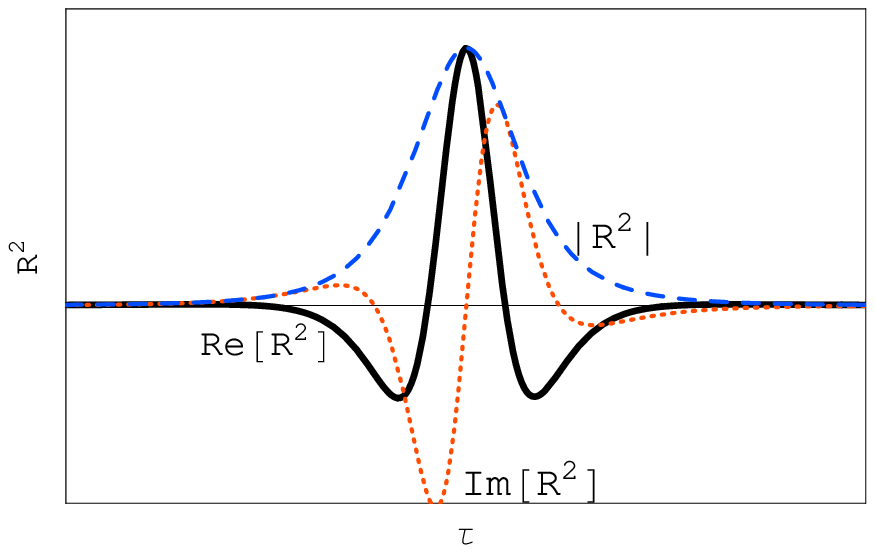}
        \label{fig:Rsquared1}
    }
    \subfigure[]{
        \includegraphics[width=\columnwidth]{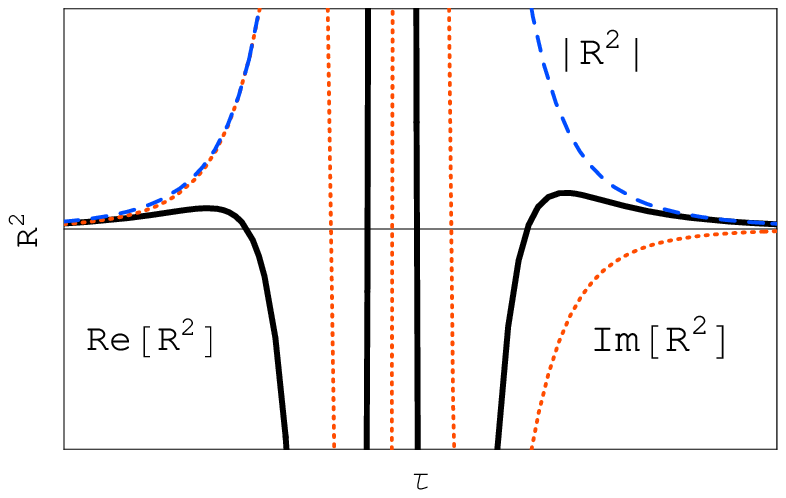}
        \label{fig:Rsquared1A}
    }
    \label{fig:Rsquared}
    \caption{The real and imaginary part and the absolute value of the
    curvature invariant (\ref{BHsingularity:hol}), which we for simplicity
    denote in figures by $R^2$.
     as a function of proper time $\tau$ for different initial
     conditions $b$.
     An infinite curvature is attained only when $\Im[b]=0$, which represents
     a set of initial conditions of measure zero.
    Fig.~\ref{fig:Rsquared1} The real and imaginary part and the absolute value of the
    curvature invariant for complex initial conditions $b$. The curvature
    invariant remains finite at all times.
    Fig.~\ref{fig:Rsquared1A} A part of the upper panel~\ref{fig:Rsquared1} zoomed in.
    At large distances the real part of the curvature invariant
    ({\it solid black line}) approaches the general relativistic
    solution, but close to and within the Schwarzschild radius
    the curvature exhibits a very different behavior, becoming even negative.
    The deviation from the general relativistic
     behavior becomes large only within the Schwarzschild
    radius and hence remains hidden behind the black hole horizon.}
\end{figure}

\begin{figure}
  \centering
  \includegraphics[width=\columnwidth]{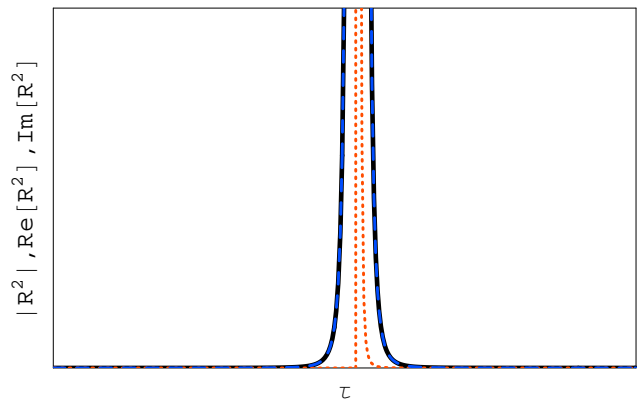}
  \caption{The real and imaginary part and the absolute value of the
    curvature invariant~(\ref{BHsingularity:hol}) with initial conditions
    corresponding to a real $b$ (vanishing initial force).
    Since $\Im[z]=0$ remains zero, the
    curvature invariant is identical to the corresponding
    curvature invariant in general relativity (\ref{BBsingularity})
    such that a curvature singularity is reached at $r = 0 = p$.}
  \label{fig:Rsquared2}
\end{figure}

 In order to verify the limited curvature conjecture, we solve~(\ref{eq:numerical})
numerically. For initial conditions, where the imaginary part of
$b$ is nonvanishing, $p$ is indeed nonzero when $r = 0$, as can be
seen from figure~\ref{fig:Radius1}. In this case there is no
curvature singularity. When $\Im[b]=0$ however, then $r$ and $p$
simultaneously go to zero, the geodesics end at a curvature
singularity, just like in the case of general relativity. This
situation is illustrated in figure~\ref{fig:Radius2}. Even though
numerical solution indicates that the evolution continues after
$r=0=p$ is reached, at that point there is a branch point of the
evolution, and the numerical integrator picks one of the Riemann
sheets (this is indicated by the cuspy feature of the numerical
solution at $r=p=0$ in figure~\ref{fig:Radius2}). Moreover, at
$r=0=p$ the curvature invariant~(\ref{BHsingularity:hol}) becomes
singular, as can be seen in figure~\ref{fig:Rsquared2}. Since
$\Im[b]=0$ only when the initial force $f_0^0=0$ exactly, the set
of initial conditions where $b$ is real is of measure zero, when
compared to the set of all initial conditions, where $b$ can be an
arbitrary complex number. Hence we conclude that the curvature
singularity is {\it not} seen by most of infalling observers.

Finally, we need to interpret what it means for $r$ to become
negative, see figure \ref{fig:Radius}. In the view of
Eq.~(\ref{eq:numerical}), for instance, we can take the sign out of
the denominator and put it in the numerator
\begin{eqnarray}
\frac{M}{- r +i (G_N/c^3)p} = \frac{-M}{ r - i (G_N/c^3)p}
\,.
\end{eqnarray}
In this way we effectively glue
the holomorphic Schwarzschild solution with a negative
$\Re[z]$ onto the anti-holomorphic solution with a positive $\Re[z]$
but with a negative mass: a white hole. This means that an observer
can fall through a black hole and emerge from a white hole with an opposite momentum
and {\it v.v.} Needless to say, this behavior is very different from general relativity,
where classically nothing can come out
once inside the event horizon. Finally, we note that the holomorphic and
anti-holomorphic solutions are in general not symmetric under time
reversal. This symmetry is broken, for example,
by the imaginary part of the curvature invariant, as can be seen in
figure~\ref{fig:Rsquared1}.

\section{Cosmology}
\label{Cosmology}

\noindent Since our Universe is isotropic and homogeneous on large
scales, our complex generalized theory should also possess
isotropy and homogeneity such that it reduces to
the theory of general relativity correctly at low energies. Since
our complex theories are completely specified in terms of
vielbeins, an assumption concerning the vielbein modeling an
isotropic and homogeneous universe is in place here. Let us try the
following {\it Ansatz}:
\begin{eqnarray}\label{vielbein:cosmology}
    e^a_\mu (z)
    =
    a(z_c^0) \delta^a_\mu
    \phantom{hallo}
    e^a_{\bar{\mu}} (\bar{z})
    =
    \bar{a}\big({z_c}^{\bar{0}}\big) \delta^a_{\bar{\mu}},
\end{eqnarray}
where
\begin{eqnarray}\nonumber
    a(z_c^0)
    =
    a_R(z_c^0) + i a_I(z_c^0)
    \phantom{hal}
    \bar{a}\big(z_c^{\bar{0}}\big)
    =
    a_R\big(z_c^{\bar{0}}\big) - i a_I\big(z_c^{\bar{0}}\big)
\,,
\end{eqnarray}
where $z_c^0=\eta+iG_NE_c$ denotes a conformal complex `time.'
(In this section we set the speed of light $c=1$.)
With these assumptions the holomorphic
Christoffel symbols~(\ref{eq:holomorphic Levi-Civita
connection}) are then given by
\begin{eqnarray}\label{eq:Christoffel holomorphic symbol FRW}
    \Gamma^\rho_{\mu\nu}
    =
    \frac{a'}{a}
     \left(
           \delta^0_\nu\delta^\rho_\mu
           + \delta^0_\mu\delta^\rho_\nu - \eta^{\rho 0}\eta_{\mu\nu}
      \right)
\,
\end{eqnarray}
and its complex conjugate, where $a' = {da}/{d z_c^0}$. Inserting
Eq.~(\ref{eq:Christoffel holomorphic symbol FRW})
into the holomorphic Riemann tensor~(\ref{eq:holomorphic Ricci tensor}) we
obtain
\begin{eqnarray}\nonumber
    R_{\mu\nu}
    &=&
    \left[\frac{a''}{a} - \left(\frac{a'}{a}\right)^2\right]
    \left(\eta_{\mu\nu}-2\delta^0_\nu\delta^0_\mu\right)
    \\\nonumber
    &+&
    \left(\frac{a'}{a}\right)^2[2\delta^0_\nu\delta^0_\mu + 2\eta_{\mu\nu}]
\end{eqnarray}
and its complex conjugate. The holomorphic Ricci scalars are then
given by
\begin{eqnarray}\nonumber
    R
    =
    6 \frac{a''}{a^3}
    =
    6\left[\frac{\ddot{a}}{a}
                     - \left(\frac{\dot{a}}{a}\right)^2
                  \right]
\end{eqnarray}
and its complex conjugate, where we have contracted holomorphic
Ricci tensors with the inverse metric $g^{\mu\nu} = ({1}/{a^2})
\eta^{\mu\nu}$ and its complex conjugate, and $\dot{a} =
{da}/{d z^0}$, $dz^0 = adz_c^0$.
The holomorphic Einstein's equations in four dimensions are
\begin{eqnarray}\nonumber
    R_{\mu\nu}
    =
    8\pi G_N (T_{\mu\nu} - \frac{1}{2}g_{\mu\nu}T)
\end{eqnarray}
and its complex conjugate, of course. Because of the isotropy and
homogeneity symmetries imposed, the Einstein's equations contain
only four independent equations due to homogeneity and isotropy
symmetries, namely one space-space ($ii$) equation,
\begin{eqnarray}\label{eq:ij holomorphic component}
    \frac{\ddot{a}}{a} + 2 \left(\frac{\dot{a}}{a}\right)^2
    =
   4 \pi G(\rho - p),
\end{eqnarray}
the time-time ($00$) equation,
\begin{eqnarray}\label{eq:friedmann holomorphic equations first}
    \frac{\ddot{a}}{a}
    =
    - \frac{4\pi G}{3} (\rho + 3 p)
\end{eqnarray}
and the complex conjugates of these equations. We can simplify
Eq.~(\ref{eq:ij holomorphic component}) by making use of
the time component (\ref{eq:friedmann holomorphic equations
first}) to obtain
\begin{eqnarray}\label{eq:friedmann holomorphic equations second}
    H^2 \equiv \left(\frac{\dot{a}}{a}\right)^2
    =
    \frac{8 \pi G_N}{3}\rho,
\end{eqnarray}
where $H={\dot{a}}/{a}$ denotes the  holomorphic Hubble parameter.
Equation~(\ref{eq:friedmann holomorphic
equations second}) together with equation~(\ref{eq:friedmann
holomorphic equations first}) constitute the holomorphic
Friedmann equations. It is
easy to see that the holomorphic Friedmann equations are
consistent with the holomorphic energy conservation equations
\begin{eqnarray}\label{rhodot}
    \dot{\rho} + 3\frac{\dot{a}}{a}(\rho + p) = 0
\end{eqnarray}
and its complex conjugate, which is derived from the holomorphic
energy conservation equations
\begin{eqnarray}\nonumber
    \partial_\mu T^\mu_{\phantom{\lambda}0}
    +
    \Gamma^\mu_{\mu\lambda} T^\lambda_{\phantom{\lambda}0}
    -
    \Gamma^\lambda_{\mu0}T^\mu_{\phantom{\lambda}\lambda}
    =
    0
\end{eqnarray}
and its complex conjugate, where we have used the expression of
the energy-momentum tensor (\ref{eq:energy momentum holomorphic
tens invar rho and p}) and holomorphic Christoffel symbol
(\ref{eq:Christoffel holomorphic symbol FRW}).
Making use of
Eqs.~(\ref{eq:energy momentum holomorphic tens invar rho and p}) and
(\ref{eq:stres-energy tensor holomorphic}),
we can express
the energy density  $\rho$ and pressure $p$ in terms of
the scalar field $\phi$. In an isotropic and homogeneous
universe the scalar field is a holomorphic function of $z^0$,
$\phi = \phi(z^0)$, such that we have
\begin{eqnarray}\label{eq:rho in terms of phi holomorphic}
   \rho
   =
   \frac{1}{2}\dot{\phi}^2
   +
   V(\phi)
\,,\qquad
    p
    =
    \frac{1}{2}\dot{\phi}^2
    -
    V(\phi).
\end{eqnarray}
The equations for $\bar{\rho}$ and $\bar{p}$ are just the complex
conjugates of these expressions. The equation of motion for the
scalar field (\ref{eq:scalar field holomorphic equations of
motion}) becomes
\begin{eqnarray}\label{eq:scalar field holomorphic equation of motion}
   \ddot{\phi} + 3H \dot{\phi} + \frac{d V}{d \phi}
   =
   0
\end{eqnarray}
and the holomorphic Friedmann
equations~(\ref{eq:friedmann holomorphic equations first}--\ref{eq:friedmann
holomorphic equations second})
then become
\begin{eqnarray}\label{eq:friedmann equations holomorphic first homogeneous}
    \frac{\ddot{a}}{a}
    =
    - \frac{8\pi G_N}{3} \left({\dot{\phi}}^2 - V\right)
\end{eqnarray}
and
\begin{eqnarray}\label{eq:friedmann equations holomorphic second homogeneous}
  H^2\equiv   \left(\frac{\dot{a}}{a}\right)^2
    =
    \frac{8 \pi G_N}{3}\left(\frac{\dot{\phi}^2}{2} +  V\right)
\end{eqnarray}
and the corresponding complex conjugates. Just like
in general relativity, one can show that only three equations
among Eqs.~(\ref{rhodot})
and~(\ref{eq:scalar field holomorphic equation of
motion}--\ref{eq:friedmann equations holomorphic second homogeneous})
are independent.

 We shall now study the holomorphic Friedmann equations for
different physical circumstances and compare the results with
those of general relativity.

\subsection{Power law expansion}
\label{Power law expansion} \noindent Let us now consider a
homogeneous and holomorphic fluid with the equation of state,
\begin{equation}
  w = \frac{p}{\rho}
\,,\qquad w \in {\bm C}\,,\qquad \Re[w]\geq -1
\,.
\label{holomorphic fluid}
\end{equation}
In this case Eq.~(\ref{rhodot}) is solved by,
\begin{equation}
  \rho = \frac{\rho_0}{a^{3(1+w)}}
\,.
\label{holomorphic fluid:rhp:soln}
\end{equation}
Next Eqs.~(\ref{eq:friedmann equations holomorphic first homogeneous})
and~(\ref{eq:friedmann equations holomorphic second homogeneous}) can be
combined into,
\begin{equation}
 \epsilon \equiv -\frac{\dot H}{H^2} = \frac{d}{dz^0}\left(\frac{1}{H}\right)
                = \frac{3}{2}(1+w)
\,,
\label{epsilon:eom}
\end{equation}
such that $\epsilon\in \bm C$ is a complex constant and $\Re[\epsilon]\geq 0$.

When equation~(\ref{epsilon:eom}) is integrated,
one gets the holomorphic expansion rate,
\begin{equation}
  H = \frac{1}{\epsilon z^0}
\,.
\label{H:holomorphic}
\end{equation}
This is easily integrated, to yield the holomorphic scale factor,
\begin{equation}
  a = \left(\frac{z^0}{\zeta_0}\right)^{{1}/{\epsilon}}
\,.
\label{a:holomorphic}
\end{equation}
These solutions are a generalization of
power law expansion of general relativity,
in that $\epsilon$ is in general a complex parameter.
$\zeta_0$ is an arbitrary constant parameter that signifies the complex
time at which $a=1$. Its physical significance is revealed by
realizing that $\zeta_0$  parametrizes the Hubble rate at the time $z^0=\zeta_0$,
as can be seen from Eqs.~(\ref{H:holomorphic})
and~(\ref{H2:holomorphic:constraint}).

 Finally, the last equation we need to solve
is~(\ref{eq:friedmann equations holomorphic second homogeneous}),
which gives,
\begin{eqnarray}
  H^2_0  = \frac{1}{\epsilon^2 \zeta_0^2}
         =  \frac{8\pi G_N\rho_0}{3}
\,,
\label{H2:holomorphic:constraint}
\end{eqnarray}
where $H_0 = H(\zeta_0)$, $\rho_0 = \rho(\zeta_0)$.

\bigskip

 We shall now show that, non unlike in general relativity,
the holomorphic power law expansion~(\ref{a:holomorphic}) can be
realised by a homogeneous holomorphic scalar field $\phi=\phi(z^0)$
with an exponential potential.
Just like in general relativity, the power law solution is then
realised in the scaling limit, in which case it exhibits attractor
behavior~\cite{Ratra:1987rm}.

Let us consider the following Lagrangian density
%
\begin{equation}
 {\cal L} = \frac12{\dot\phi^2} - V(\phi)\,,\qquad
V(\phi) = V_0\exp\left(-\lambda \frac{\phi}{M}\right)
\,.
\label{holomorphic Lagrangian}
\end{equation}

We now assume that the Friedmann equation permits a scaling
solution of the form, $H\propto 1/z^0$.
Obviously, such a solution must satisfy,
\begin{equation}
 \dot \phi = \dot\phi_0 \frac{\zeta_0}{z^0}
\,, \qquad
\phi = \phi_0 + \dot\phi_0\zeta_0\ln\left(\frac{z^0}{\zeta_0}\right)
\,,
\label{dotphi:phi}
\end{equation}
where $\phi_0, \dot\phi_0$ and $\zeta_0$ are (complex) constants
(note that $\zeta_0$ and $\phi_0$ are not independent; indeed,
rescaling $\zeta_0$ can be compensated by the appropriate shift in $\phi_0$).
Note that Eq.~(\ref{a:holomorphic}) implies that the scalar
field~(\ref{dotphi:phi}) can be considered as a `marker' for
the scale factor,
\begin{equation}
   a = \exp\left(\frac{\phi-\phi_0}{\epsilon\dot\phi_0\zeta_0}\right)
\label{a-phi:connection}
\end{equation}
and in this sense defines a clock.

A constant $w$ requires the scaling of the
potential~(\ref{holomorphic Lagrangian}),
\begin{equation}
  V = V_0 \exp\left(-\frac{\lambda\phi_0}{M}\right)
              \left(\frac{\zeta_0}{z^0}\right)
                           ^{(\lambda\dot\phi_0\zeta_0)/M}
\,.
\end{equation}
We are free to absorb $\phi_0$ in $V_0$ by redefining,
$V_0\exp\left(-{\lambda\phi_0}/{M}\right)\rightarrow V_0$.
Demanding the scaling,
\begin{equation}
  V = V_0 \left(\frac{\zeta_0}{z^0}\right)^2
\,,
\label{V:attractor}
\end{equation}
implies the following condition on the initial field velocity,
\begin{equation}
 \dot\phi_0 = \frac{2M}{\lambda \zeta_0}
\,,
\label{dotphi:attractor}
\end{equation}
where $V_0 = V(\phi_0)=V(\phi(\zeta_0))$.~\footnote{
One can show that, even when the condition~(\ref{dotphi:attractor})
is not met,
quite generally $\dot\phi$ eventually approaches the attractor
solution for which~(\ref{dotphi:attractor}) holds.}

Taking account of these scalings, the equation of state parameter $w$
becomes indeed constant,
\begin{equation}
  w = \frac{\eta_0-1}{\eta_0+1}
\,,\qquad  \eta_0 = \frac{\dot\phi_0^2}{2V_0}
                  = \frac{2M^2}{\lambda^2 V_0\zeta_0^2}
\label{w:scalar}
\end{equation}
and hence $\epsilon$ in~(\ref{epsilon:eom}) is also constant,
\begin{equation}
 \epsilon = \frac{d}{dz^0}\left(\frac{1}{H}\right)
                = \frac{3\eta_0}{1+\eta_0}
\,.
\label{epsilon:eom:2}
\end{equation}
The conservation equation is now trivially satisfied,
implying the energy density scaling,
$\rho = \rho_0/a^{2\epsilon} = \rho_0(\zeta_0/z^0)^2$.
Finally, the Friedmann
equation~(\ref{eq:friedmann equations holomorphic second homogeneous})
and~(\ref{H2:holomorphic:constraint}) yields the constraint
\begin{equation}
    H^2_0  = \frac{8 \pi G_N V_0}{3-\epsilon}
\,.
\label{Holomorhic:Friedmann:2}
\end{equation}
When this is combined with Eq.~(\ref{epsilon:eom:2}),
one gets the following algebraic equation for $\epsilon$
\begin{equation}
 \epsilon\left(\epsilon-3\right)
 \left(\epsilon-\frac{\lambda^2}{16\pi G_NM^2}\right)
 = 0
\,,
\label{epsilon:constraint}
\end{equation}
which links the parameters of the scalar theory with
$\epsilon$. Note that $\epsilon$ does not depend on
the initial conditions on the field, but only on the coupling
parameters of the potential.
This should not surprise us since we are considering an attractor solution.

From Eq.~(\ref{epsilon:constraint}) it follows that
de Sitter space ($\epsilon=0$) and kination ($\epsilon=3)$ are solutions
and have a special significance.
One can show that $\Re[\epsilon]\leq 3$, the limit is saturated when
$\lambda^2 = 16\pi G_NM^2$. For larger values of $\lambda$
there is no scaling solution: $V$ redshifts faster than the kinetic term,
resulting in $\epsilon\rightarrow 3$.
de Sitter space $\epsilon=0$ is also a solution of
Eq.~(\ref{epsilon:constraint}), which is realised when $\dot\phi_0=0$.
This is indeed a stable solution, for which
kinetic energy vanishes, and $V=V_0 \equiv \Lambda_0/(8\pi G_N)$,
where $\Lambda_0$ denotes the equivalent cosmological constant,
which in holomorphic gravity can be complex.

 The nontrivial solution of Eq.~(\ref{epsilon:constraint}) is
\begin{equation}
\epsilon=\frac{\lambda^2}{16\pi G_NM^2}
\,,\qquad
 0\leq  \Re[\epsilon]\leq 3
\label{epsilon:constraint:2}
\end{equation}
which corresponds to the attractor solution when
the condition $ 0\leq  \Re[\epsilon]\leq 3$
 in Eq.~(\ref{epsilon:constraint:2}) is satisfied.
Of course, the solution~(\ref{epsilon:constraint:2}) is more general
than the corresponding solution of general relativity in that
$\lambda$ and hence also $\epsilon$ can be complex.
To appreciate the significance of a complex $\epsilon$, let us
assume that the Universe follows the attractor behavior with
$\epsilon$ given by~(\ref{epsilon:constraint:2}). In this case the
exponential potential~(\ref{holomorphic Lagrangian})
can be written as~\footnote{Note that the scalar
potential of hermitian gravity discussed
in Ref.~\cite{MantzHermitianGravity} differs from
Eq.~(\ref{attractor potential}).},
\begin{equation}
  V \;\; \stackrel{\rm attractor}{\longrightarrow}
    \;\; V_0 \exp\left(-\sqrt{16\pi G_N\epsilon}\;\phi\right)
\,,\qquad
 0\leq  \Re[\epsilon]\leq 3
\,.
\label{attractor potential}
\end{equation}
This potential can be used to obtain both an accelerating universe
(for which $\Re[\epsilon]<1$) and a decelerating universe
(with $\Re[\epsilon]>1$).
Thus with the appropriate choice of $\lambda$
all standard cases in cosmology can be reproduced:
radiation era ($\epsilon =2$);
matter era ($\epsilon =3/2$); kination ($\epsilon =3$), which is
realized in the limit when $\lambda\rightarrow 0$ and $V_0\rightarrow 0$;
inflation ($0<\Re[\epsilon]\ll 1 $), {\it etc}.

The physical Hubble parameter ${\cal H}$ is given
in terms of the real part of the holomorphic expansion rate $H$ as,
${\cal H}^2=\Re [H^2]$.
For a real $\epsilon$ Eq.~(\ref{H:holomorphic}) implies,
\begin{eqnarray}\label{eq:Hubble hol:physical:2}
    \Re[H^2] = \frac{t^2-G_N^2E^2}
                    {\epsilon^2 [t^2+G_N^2E^2]^2}
\,.
\end{eqnarray}
For a complex $\epsilon$ the expression is more complicated,
\begin{equation}\label{eq:Hubble hol:physical:3}
    \Re[H^2] = \frac{(\epsilon_R^2-\epsilon_I^2)[t^2-G_N^2E^2]
                     -4\epsilon_R\epsilon_I tG_NE }
                    {|\epsilon|^4 [t^2+G_N^2E^2]^2}
.
\end{equation}

 In order to find out whether the holomorphic Big Bang singularity
is ever reached, we need to study geodesic equations, which
shall give us a crucial information on whether
a freely falling observer experiences the holomorphic singularity
in~(\ref{eq:Hubble hol:physical:2}--\ref{eq:Hubble hol:physical:3})
({\it cf.} Ref.~\cite{MantzHermitianGravity}).

The physical scale factor ${\cal A}$ can be obtained as the
real part of the holomorphic scale factor ~(\ref{a:holomorphic}) squared,
\begin{eqnarray}\nonumber
 {\cal A}^2 = \left(\frac{t^2 + G_N^2E^2}{|\zeta_0|^2}
              \right)^{\frac{\epsilon_R}{|\epsilon|^2}
                       -\frac{\epsilon_I}{|\epsilon|^2}
                             {\rm Arctan}\left(\frac{G_NE}{|t|}\right)}
\times
\\\nonumber
\times\cos\Big(\left[\frac{2\epsilon_I}{|\epsilon|^2}
               +\frac{2\epsilon_R}{|\epsilon|^2}
                     {\rm Arctan}\left(\frac{G_NE}{|t|}\right)\right]
\times
\\\label{scale factor:physical}
\times
               \ln\left(\frac{t^2 + G_N^2E^2}{|\zeta_0|^2}\right)\Big)
\,,
\end{eqnarray}
where $\epsilon = \epsilon_R + i \epsilon_I$.
When $\epsilon_I=0$ and provided $E$ does not grow with time
(which is reasonable), the Universe approaches the standard FLRW cosmology.
When however $\epsilon_I\neq 0$, the Universe's scale factor
develops oscillations, which can result in significant differences
between holomorphic and FLRW cosmology even at late times.
This is a disadvantage of holomorphic cosmology when compared with,
for example, Hermitian cosmology developed
in Ref.~\cite{MantzHermitianGravity}.
Note that, when $\epsilon$ develops an imaginary part, then the
potential~(\ref{attractor potential}) violates charge-parity symmetry.
In this case, as $\phi$ evolves, $\Re[V]$ oscillates and
can be either positive or negative. When the potential is negative,
the Universe can enter an anti-de Sitter-like phase.
As a consequence, the physical
scale factor ${\cal A}^2$ in Eq.~(\ref{scale factor:physical}) can be
either positive or negative. To prevent a negative value for
${\cal A}^2$ one can add a constant to $V$ (or a cosmological term),
which will keep ${\cal A}^2$ positive at all times.
This type of behavior can have relevance for the Universe's dark energy.



\subsection{Geodesic equation}
\label{Geodesic equation}

 In order to better understand the behavior of the expansion rate
and the corresponding scale factor, we shall now consider a
freely falling observer in the contracting phase.
To do that, we need to solve the corresponding geodesic equation,
which in holomorphic gravity has formally the same form as
the corresponding geodesic equation of general relativity
discussed for example in Ref.~\cite{MantzHermitianGravity}.
Taking account of the Christoffel
symbol~(\ref{eq:Christoffel holomorphic symbol FRW}),
the geodesic equation and the line element can be written in conformal time as,
\begin{eqnarray}\label{FLRW:geodesic eq}
  \frac{du_c^\mu}{d\tau}
    + \frac{a'}{a}\bigg(2u_c^0u_c^\mu
             - \frac{\delta^\mu_{\,0}}{a^2}\bigg)
   &=&
   0,
 \\\nonumber
   a^2\eta_{\alpha\beta}u_c^\alpha u_c^\beta
   + \bar a^2\eta_{\alpha\beta}u_c^{\bar\alpha} u_c^{\bar\beta}
   &=&
   -2
\end{eqnarray}
where $\tau$ is the (real) proper time of a freely falling observer
(in the frame in which all 3-velocities vanish): $(ds)^2=-2(d\tau)^2$,
and
\begin{equation}
 u^\mu_c =\frac{dx_c^\mu}{d\tau}
\label{ucmu}
\end{equation}
is the 4-velocity in conformal coordinates $x_c^\mu = (\eta,x^i_c)$.
Defining the physical 4-velocity as
\begin{equation}
 u^\mu =au_c^\mu = a\frac{dx_c^\mu}{d\tau}
\label{umu}
\end{equation}
we can rewrite the spatial and time component of
the geodesic equation~(\ref{FLRW:geodesic eq}) as,
\begin{equation}
  \frac{d (au^i)}{d\tau} = 0
\,,\qquad    \frac{d \{a^2[(u^0)^2-1]\}}{d\tau} = 0
\,.
\label{FLRW:geodesic eq:i+0}
\end{equation}
These are solved by the following scaling solution,
\begin{equation}
  \frac{(u^0)^2-1}{(u_0^0)^2-1} = \frac{a_0^2}{a^2}
\,,\qquad
   \frac{u^i}{u_0^i} = \frac{a_0}{a}
\,,
\label{FLRW:geodesic eq:her:sol:1}
\end{equation}
where  $u_0^\mu=u^\mu(\tau_0)$ and $u^\mu = u^\mu(\tau)$.
We introduce a complex constant $U$
by recasting the first equation in~(\ref{FLRW:geodesic eq:her:sol:1})
as
\begin{eqnarray}
  u^0 &\equiv& \frac{dz^0}{d\tau} = \pm\sqrt{\frac{U}{a^2}+1}
\nonumber
\\
   U &\equiv& a_0^2[(u_0^0)^2-1]  = U_R + i U_I
\,.
\label{FLRW:geodesic eq:her:sol:2}
\end{eqnarray}
When the line element~(\ref{FLRW:geodesic eq})
is taken account of,
one would be tempted to identify,
\begin{equation}
U=a_0^2\|\vec u_0\|^2
\,.
\label{Uvsvecu}
\end{equation}
From the line element~(\ref{FLRW:geodesic eq}) it follows that,
strictly speaking, only the real part of Eq.~(\ref{FLRW:geodesic eq:her:sol:2})
must be satisfied. Yet by holomorphy (which we assume to dictate
a unique analytic extension) we know that also the imaginary parts must match.
We assume here that Eq.~(\ref{FLRW:geodesic eq:her:sol:2}) indeed holds true.

Upon rewriting Eq.~(\ref{FLRW:geodesic eq:her:sol:2}) in the integral form
we get,
\begin{equation}
 \int \frac{da a^\epsilon}{\sqrt{a^2+U}} = \frac{\tau}{\epsilon\zeta_0}
\,,
\label{a-eta:integral}
\end{equation}
where we made use of Eq.~(\ref{a:holomorphic})  and $dz^0 =
\epsilon z^0 da/a = \epsilon \zeta_0 a^{\epsilon-1} da$.
Integrating~(\ref{a-eta:integral}) yields a hypergeometric
function. Rather than performing a general analysis of
Eq.~(\ref{a-eta:integral}), we shall consider the cosmologically
interesting cases which are simple to analyze. We shall first
consider a {\it curvature dominated epoch} with $\epsilon=1$.

\subsubsection{Curvature Dominated Epoch}
\noindent In this case Eq.~(\ref{a-eta:integral}) integrates to,
\begin{equation}
 \sqrt{a^2+U} = \frac{\tau}{\zeta_0}
\,.
\label{a-eta:curvature-dom}
\end{equation}
Note that that is not a unique expression; there is a freedom to shift
$\tau$ for an arbitrary real constant.
From Eq.~(\ref{a-eta:curvature-dom}) it follows that,
\begin{eqnarray}\nonumber
 a &=& \sqrt{\left(\frac{\tau}{\zeta_0}\right)^2 - U}
\\
  z^0 &=& \sqrt{\tau^2 - U\zeta_0^2}
      =\sqrt{\tau^2 - \zeta_0^2\|\vec u\|_0^2}
\,,
\label{a-eta:curvature-dom:2}
\end{eqnarray}
from where we conclude,
\begin{equation}
 H^2 = \frac{1}{(z^0)^2}
  = \frac{1}{\tau^2 - \zeta_0^2\|\vec u\|_0^2}
\,.
\label{H:curvature-dom}
\end{equation}
\begin{figure}[ht]
  \centering
    \subfigure[]{
        \includegraphics[width=\columnwidth]{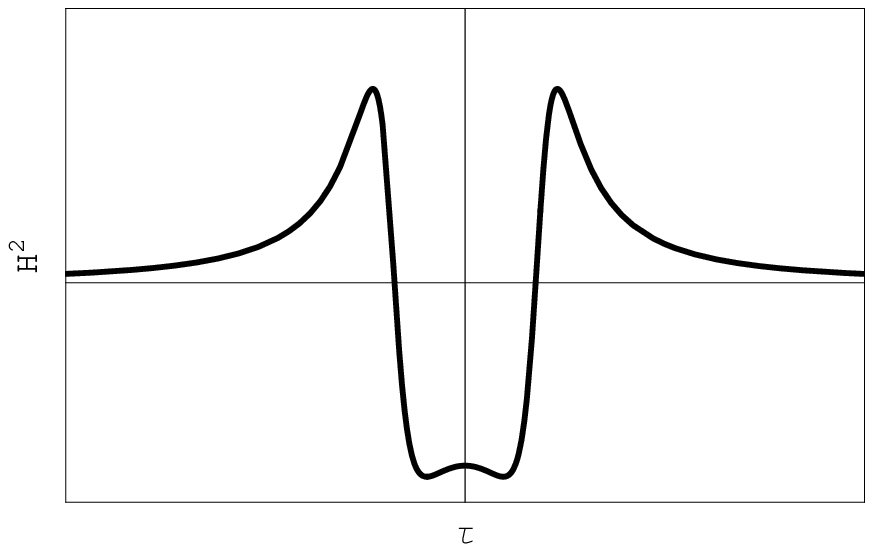}
        \label{fig:Hubble1A}
    }
    \subfigure[]{
        \includegraphics[width=\columnwidth]{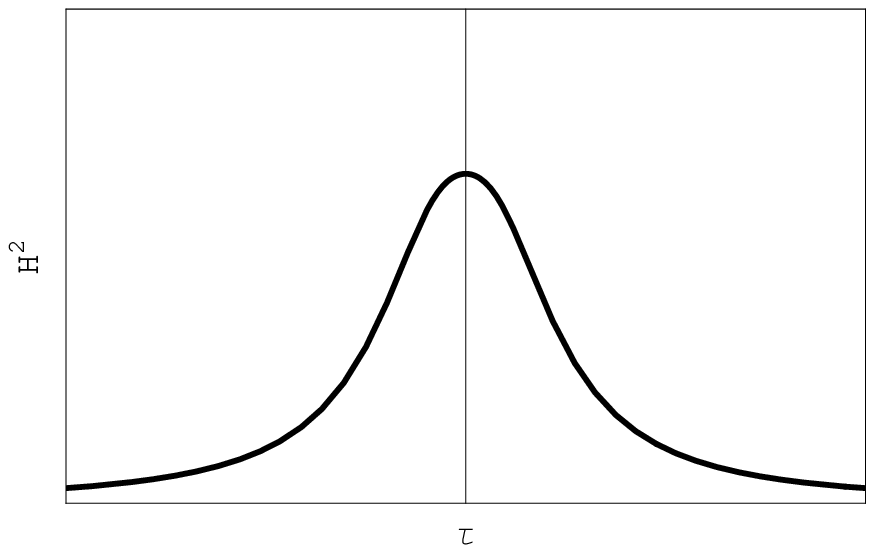}
        \label{fig:Hubble1B}
    }
    \label{fig:Hubble1}
    \caption{The physical Hubble parameter (\ref{H:curvature-dom:2})
    in the curvature dominated epoch as a function of the proper time for different values of
    $\vartheta$ remains finite at all times. An infinite
    value of the physical
    Hubble parameter is attained when $\Im[\vec u_0]=0$,
    which represents a set of initial conditions
    of measure zero compared to the set of all initial conditions.
    \ref{fig:Hubble1A} The physical Hubble parameter with initial conditions
      $\text{cos}(\vartheta) < 0$.
    \ref{fig:Hubble1B} The physical Hubble parameter with initial conditions
     $\text{cos}(\vartheta) > 0$.}
\end{figure}
From~(\ref{a-eta:curvature-dom:2}) we get for the physical scale
factor,
\begin{eqnarray}\nonumber
{\cal A}^2 &=& \Re[a^2] =
\frac{\tau^2}{|\zeta_0|^2}\cos(2\theta_\zeta)
                      - |\vec u|_0^2\cos(2\theta_u)
\\
 \|\vec u\|_0 &=& |\vec u|_0{\rm e}^{i\theta_u}
 \,,\quad
 \zeta_0=|\zeta_0|{\rm e}^{i\theta_\zeta}
 \,,
\label{a-eta:curvature-dom:3}
\end{eqnarray}
This then implies that the scale factor ${\cal A}^2$ will
be positive at all times, provided both $\cos(2\theta_\zeta)>0$ and
$\cos(2\theta_u)<0$, in which case the Big Bang singularity
will never be reached. In order to have a more
precise understanding on whether the Big Bang
singularity is ever attained, we need to study
a curvature invariant, one example being
the Hubble parameter.

 Observe that the physical Hubble parameter, which is obtained
from Eq.~(\ref{H:curvature-dom}),
\begin{equation}
{\cal H}^2 = \Re[H^2]
  = \frac{\tau^2 - |\vec u|_0^2 |\zeta_0|^2\cos(\vartheta)}
         {(\tau^2 - |\vec u|_0^2 |\zeta_0|^2\cos(\vartheta))^2
          +|\vec u|_0^4 |\zeta_0|^4\sin^2(\vartheta)}
\,.
\label{H:curvature-dom:2}
\end{equation}
represents a {\it bounce universe}, where we defined
   $ \vartheta
    =
    2(\theta_u+\theta_\zeta)$.
When $\cos(\vartheta)>0$ the maximal physical expansion rate is
reached when $\tau^2=|\vec u|_0^2 |\zeta_0|^2\cos(\vartheta)
       + |\vec u|_0^2 |\zeta_0|^2\sin(\vartheta)$,
for which
\begin{equation}
{\cal H}_{\rm max}^2
        = \frac{1}{2|\vec u|_0^2 |\zeta_0|^2\sin(\vartheta)}
\,.
\label{H:curvature-dom:3}
\end{equation}
%
%

As can be seen in figure \ref{fig:Hubble1A}, this corresponds to a
local maximum. At $\tau^2=|\vec u|_0^2
|\zeta_0|^2\cos(\vartheta)$,
 ${\cal H}=0$; at even smaller proper times ${\cal H}^2<0$,
which means that a local observer will have the impression
that the Universe has entered an anti-de Sitter-like phase.
Since we are in holomorphic gravity,
there is no need to change the form of the line element (vielbein)
{\it Ansatz}~(\ref{vielbein:cosmology}).
The minimal expansion rate squared ${\cal H}_{\rm min}^2$
is reached when $\tau=0$,
\begin{equation}
{\cal H}_{\rm min}^2
        = -\frac{\cos(\vartheta)}
                {|\vec u|_0^2 |\zeta_0|^2}
\,,
\label{H:curvature-dom:4}
\end{equation}
which is singular only when $\vec u_0=0$, or equivalently when $u_0=1$,
corresponding to a set of initial conditions of measure {\it zero}.

It is interesting to note that when $\cos(\vartheta)<0$, the
expansion rate~(\ref{H:curvature-dom:4}) becomes a global maximum,
away from each ${\cal H}$ decreases monotonously in both
directions, as can be seen in figure \ref{fig:Hubble1B}. This case
represents a more conventional bounce Universe, and it is realized
when the initial 3-force dominates over the initial 3-velocity,
$G_N||\vec f(\tau_0)\,||>||\vec u(\tau_0)\,||$.

\subsubsection{Radiation Era}

\begin{figure}[!]
    \subfigure[]{
        \includegraphics[width=\columnwidth]{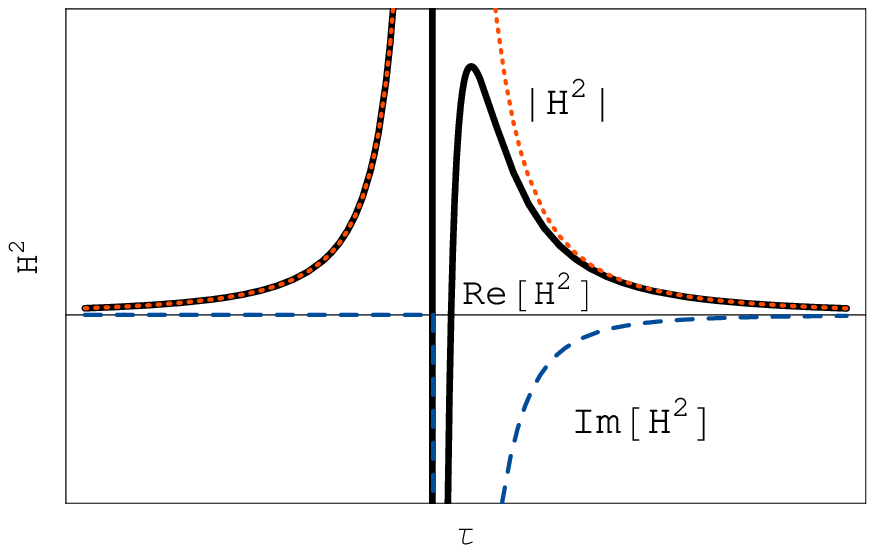}
        \label{fig:Hubbe2C}
    }
    \subfigure[]{
        \includegraphics[width=\columnwidth]{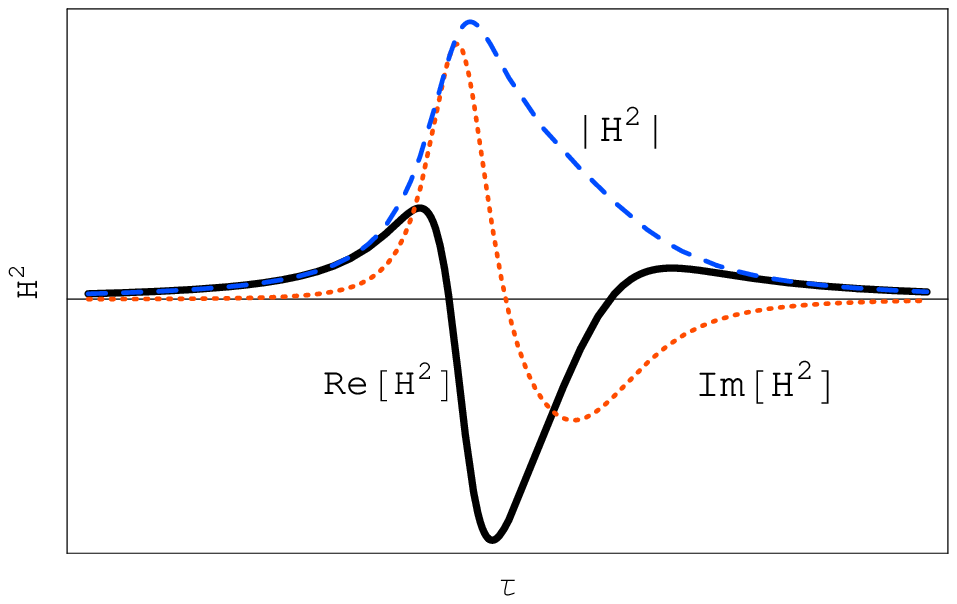}
        \label{fig:Hubble2B}
    }
    \label{fig:Hubble2}
    \caption{The physical Hubble parameter in radiation era
    $(\epsilon = 2)$
    as a function of proper time $\tau$ for different initial conditions.
     \ref{fig:Hubbe2C} The physical Hubble parameter for observers
      with real initial condition $U$
      defined in Eq.~(\ref{FLRW:geodesic eq:her:sol:2}) becomes singular
      at some time $\tau$, which corresponds to the Big Bing singularity.
     \ref{fig:Hubble2B} When $U$ in Eq.~(\ref{FLRW:geodesic eq:her:sol:2}) is complex,
      the physical Hubble parameter remains finite at all times, thus resolving
      the Big Bang singularity.
      The Hubble parameter is in general nonsymmetric under time reversal.
      Only when $U$ is purely imaginary,
      the physical Hubble parameter
     \textit{is} symmetric under time reversal.}
\end{figure}

\noindent Let us now consider radiation era $(\epsilon =2)$, in
which case Eq.~(\ref{a-eta:integral}) integrates to,
\begin{equation}
 a\sqrt{a^2+U}
 -U\ln\left(a+\sqrt{a^2+U}\right)
 = \frac{\tau}{\zeta_0} \phantom{ha} ({\rm radiation\;era})
\,.
\label{a:radiation:1}
\end{equation}
This transcendental equation cannot be solved for $a=a(\tau)$ in
terms of elementary functions and thus it is hard to analyze in
complete generality. A rather conclusive analysis can be,
nevertheless, performed by integrating Eq.~(\ref{FLRW:geodesic eq:her:sol:2})
numerically, which shows that, when $U$ and $\zeta_0$ are chosen
real, the Hubble expansion rate
$H^2 = 1/(2z^0)^2=1/(2\zeta_0 a^2)^2$ becomes singular
at $\tau=0$, as can be clearly seen in figure \ref{fig:Hubbe2C}.
When $U = (u^0)^2_0-1=\|\vec u\,\|^2_0$ is complex however, then
-- as can be seen in figure~\ref{fig:Hubble2B} -- both $\Re[H^2]$
and $\Im[H^2]$ are in general nonzero,
implying that the physical Hubble parameter remains finite at all times.
We conclude that, just like in the case of curvature domination, there is no
curvature singularity except for a set of initial conditions of
measure zero ($\Im[u^0_0]=0$),
when compared to the set of all initial conditions ($\Im[u^0_0]$ arbitrary).  The
Big Bang singularity is hence resolved also in the radiation era.
We think that an analogous conclusion holds for more general expanding space-times.
Even though the physics of radiation era is quite different from
that of Schwarzschild black holes, the corresponding geodesic
equations~(\ref{eq:numerical}) and~(\ref{FLRW:geodesic eq:her:sol:2}) --
based on which we performed the analyses
of singularities -- are of an identical form, and thus the conclusions of the analyses
are quite similar.

\section{Conclusions}

 We have shown that -- just like Hermitian
gravity~\cite{MantzHermitianGravity} -- holomorphic gravity also
resolves the cosmological Big Bang singularity.

Furthermore, we have shown that holomorphic gravity resolves the
Schwarzschild space-time singularity. Given the similarities
between the Hermitian and holomorphic gravity theories, we expect
that Hermitian gravity also does not suffer from the singularity
problems of general relativity~\cite{MantzProkopec:2008}.

Even though, strictly speaking, singularities in complex theories of
gravity exist, they occupy just one point on an 8 dimensional phase space,
and consequently they are reached only by special geodesics/observers.
Indeed, we have argued that the space of initial conditions
of observers which encounter singularities is of measure zero,
when compared to the space of all initial conditions.
Hence quite generically observers will not see singularities
within holomorphic gravity.
We expect that quantization of
space-time and momentum-energy will lead to further smearing of
these singularities.

%
%

\section*{Acknowledgements}
\noindent
The authors acknowledge financial support by FOM grant 07PR2522
and by Utrecht University.

\bibliographystyle{apsrev}

%
%

\end{document}